\documentstyle[psfig,subfigure]{mn}
\begin{document}
\title[{\em XMM-Newton\/} Surveys of the CFRS Fields - II]{{\em
XMM-Newton\/} Surveys of the Canada-France Redshift Survey Fields -
II: The X-ray Catalogues, the Properties of the Host Galaxies, and the
Redshift Distribution}
\author[Waskett et al.]
{T. J. Waskett$^1$\thanks{E-mail: Tim.Waskett@astro.cf.ac.uk}, 
S. A. Eales$^1$,
W. K. Gear$^1$,
H. J. McCracken$^2$, \cr
M. Brodwin$^3$,
K. Nandra$^4$,
E. S. Laird$^4$,
S. Lilly$^5$ \\ 
$^1$Department of Physics and Astronomy, University of Wales Cardiff,
PO Box 913, Cardiff, CF24 3YB, UK\\ 
$^2$University of Bologna, Department of Astronomy, via Ranzani 1,
40127 Bologna, Italy\\
$^3$Department of Astronomy \& Astrophysics, University of Toronto, 60
St. George St., Toronto, Ontario, M5S 3H8, Canada\\
$^4$Astrophysics Group, Imperial College London, Blackett
Laboratory, Prince Consort Road, London, SW7 2AZ, UK\\
$^5$Herzberg Institute for Astrophysics, Dominion Astronomical
Observatory, National Research Council, Canada}

\maketitle

\begin{abstract}
We present the X-ray source catalogues for the {\em XMM\/} surveys of
the 3-h and 14-h Canada-France Redshift Survey fields ($0.5-10~keV$
flux range $\sim2\times 10^{-15} - 10^{-13}~erg~cm^{-2}~s^{-1}$).  We use a
subset of the {\em XMM\/} sources, which have {\em Chandra\/}
positions, to determine the best method of obtaining optical
identifications of sources with only {\em XMM\/} positions.  We find
optical identifications for $79~per~cent$ of the {\em XMM\/} sources
for which there are deep optical images.  The sources without optical
identifications are likely to be optically fainter and have higher
redshifts than the sources with identifications.  We have estimated
`photometric redshifts' for the identified sources, calibrating our
method using $\sim 200$ galaxies in the fields with spectroscopic
redshifts.  We find that the redshift distribution has a strong peak
at $z\sim0.7$. 

The host galaxies of AGN identified in this work cover a wide range of
optical properties with every galaxy type being represented, and no
obvious preference for one type over another.  Redder types tend to be
more luminous than blue types, particularly at lower redshifts.  The
host galaxies also span a wide range of optical luminosity, in
contrast to the narrow range found for the starburst galaxies detected
in $\mu Jy$ radio surveys.  We find a strong correlation between
optical and X-ray luminosity similar to the Magorrian relation,
although selection effects cannot be ruled out.
\end{abstract}


\section{Introduction}
Deep exposures with the most recent and powerful X-ray observatories,
{\em XMM-Newton\/} and {\em Chandra\/} (e.g. Barger et al. 2003;
Giacconi et al. 2002; Mainieri et al. 2002; M$^{c}$Hardy et al. 2003;
Page et al. 2003), have built on the deepest {\em ROSAT\/} X-ray
surveys (e.g. M$^{c}$Hardy et al. 1998; Hasinger et al. 1998) by going
deeper and to higher X-ray energies with better positional accuracy.
This has opened up the study of faint X-ray sources such as high
redshift AGN, and has also revealed X-ray emission from otherwise
normal galaxies at more modest redshifts (Hornschemeier et al. 2003).
These surveys have now resolved the majority of the cosmic X-ray
background (XRB) in the soft ($0.5-2~keV$) X-ray band with a small
fraction left unaccounted for in the hard ($2-10~keV$) band (Moretti
et al. 2003).  

The nature of the XRB at these X-ray energies is well on the way to
being understood but the peak in the XRB lies at a much higher energy
($\sim 30~keV$).  This indicates that a population of very faint
sources, with very hard spectra, make up the remaining fraction of the
XRB in the hard band, and would also be expected to contribute a much
greater fraction to the XRB nearer its peak (Moretti et al. 2003).
Such hard sources are most likely a result of extremely high
obscuration, which progressively wipes out X-ray emission from low to
high energy, turning an intrinsically soft spectrum into a much harder
observed one.

The radiation absorbed during this process must be re-emitted at
longer wavelengths and the possibility of the Far-IR/Sub-mm background
being somehow connected with the XRB is discussed in many papers
(e.g. Almaini, Lawrence \& Boyle 1999).  However, current X-ray/Sub-mm
surveys suggest that the two backgrounds are only loosely related
(e.g. Waskett et al. 2003; Alexander et al. 2003; Severgnini et
al. 2000).  Future instrumentation with higher energy limits are
likely required to fully explain the XRB and the nature of the sources
that dominate its peak.

At present though, the emphasis must be turned to those sources that
we can observe easily with the current instrumentation.  QSOs and
type-I AGN dominate the softest X-ray energies with an increasing
contribution from more obscured type-II AGN becoming important at
higher energies (e.g. Gilli, Salvati \& Hasinger 2001).  Identifying
the optical counterparts to these sources is crucial for a full
understanding of their properties and a great deal of effort has been
expended in obtaining this information (e.g. Barger et al. 2003;
M$^{c}$Hardy et al. 2003).

For example, one of the most useful quantities that can be derived
from a source list is the luminosity function.  This reveals much
about the nature of a population and determining its evolution with
redshift can shed light on how the population as a whole changes over
time.  The X-ray luminosity function (XLF) has begun to be
investigated in depth by several groups (Cowie et al. 2003; Steffen et
al. 2003; Ueda et al. 2003).  Both Ueda et al. (2003) and Steffen et
al. (2003) find that the evolution of the XLF is a function of
luminosity.  The population of X-ray sources with
$L_{X}(2-10~keV)>3\times10^{43}~erg~s^{-1}$ is dominated by type-I
AGN, and the number-density of these sources increases with redshift
out to $z\sim2-3$.  At lower X-ray luminosities however, the fraction
of type-II AGN increases rapidly with decreasing X-ray luminosity.
The number-density of these sources appears to peak at $z<1$.

Although {\em Chandra\/} is better suited for identifying X-ray
sources with optical counterparts ({\em XMM\/} has a resolution of
$\sim 6\arcsec$ full width half maximum (FWHM) cf. $\sim 0.5\arcsec$
for {\em Chandra\/}), {\em XMM\/} has greater sensitivity and a larger
field of view (FoV), making it better for large area surveys.  In this
paper we report the results of a medium-deep {\em XMM\/} survey
composed of two separate exposures ($\sim 0.4$ square degrees).  We
quantify the ability of such a survey to identify X-ray sources with
optical counterparts by comparing the IDs for a subset of the {\em
XMM\/} sources with the IDs obtained using {\em Chandra\/} positions
for the same sources.  We estimate redshifts for our identified
sources using photometric redshift codes.  These allow a quick, and
reasonably reliable, way of obtaining redshifts for objects with
multi-band photometry.  Although not as accurate as spectroscopy these
techniques are becoming widely used as a short-cut for large surveys,
where statistical properties are fairly insensitive to the accuracy of
individual redshift measurements (Csabai et al. 2003; Fontana et
al. 2000; Kashikawa et al. 2003).  These methods can also be used on
objects fainter than the spectroscopic limit, where many X-ray source
counterparts reside (Alexander et al. 2001).  We test two photometric
redshift estimation codes on our X-ray source IDs and obtain a robust
redshift distribution for those sources that could be identified
reliably, while placing limits on the properties of those that could
not.

Ultimately we will use our identified AGN, and their redshifts, to
construct the XLF for different populations, and calculate its
evolution with redshift.  The results of this study will be reported
in paper-III, the next in this series.

We assume an $H_{0}$ of $75~km~s^{-1}~Mpc^{-1}$ and a concordance Universe 
with $\Omega_{M}=0.3$ and $\Omega_{\Lambda}=0.7$.

\section{X-ray Data}
Two {\em XMM\/} surveys are considered in this work, X-ray surveys of
the CFRS 3 and 14-h (also known as the Groth Strip) fields (Lilly et
al. 1995a).  The data reduction for these surveys, together with the
comparison between SCUBA and {\em XMM\/} data, are described in detail
in Waskett et al. (2003) (paper-I).  The 14-h XMM data was first
presented in Miyaji \& Griffiths (2001).  Both surveys are of $\sim
50~ks$ duration.  In this section we summarise some of the key points
of the data analysis.

The raw X-ray data were reduced using v5.3 of the SAS software for
{\em XMM\/}.  Because {\em XMM\/} has a large spectral range the data
were divided into two energy bands; the soft band includes photons in
the range $0.5-2~keV$ while the hard band covers $2-10~keV$.  {\em
XMM\/} has three X-ray cameras that operate simultaneously, so in
total six images were used for the source detection: soft and hard
bands for each of the two MOS cameras and also the PN camera.  The
source detection was performed simultaneously on all six images using
the sliding box and maximum likelihood detection procedure within the
SAS software, with the source extent fitting turned on.  A photon
index $\Gamma=1.7$ was assumed for the counts to flux conversion in
both bands.  The thresholds for the source detection were set to 10
for the sliding box part and 15 for the maximum likelihood part,
ensuring sources were only detected at greater than about $4~\sigma$
above the local background.  Final source parameters were derived
using data from both bands and all three instruments, for maximum
accuracy and to minimise spurious detections from any single camera,
while probing fainter fluxes.  Using both soft and hard bands
simultaneously also allows the detection routine to calculate a more
accurate full $0.5-10~keV$ flux.  The final source list contains,
amongst other parameters: source positions, fluxes in the soft, hard
and full combined bands and the vignetting corrected hardness ratios
for each source. 

For this work the hardness ratio is defined as: $$HR = {{N(H) - N(S)}
\over {N(H) + N(S)}}$$ where $N(H)$ and $N(S)$ are the counts observed
for a source in the hard and soft bands respectively, after correction
for vignetting.  Higher values indicate a harder spectrum.

In total there are 146 sources detected in the 3-h field and 154 in
the 14-h field.  Most are point sources.  Tables~\ref{3-hXray}
\&~\ref{14-hXray} list the basic properties of a sample of the X-ray
sources, in the two fields; the full tables for all the sources appear
in the electronic version of the paper.  Throughout this paper sources
labelled with 3.* refer to 3-h field sources and those labelled with
14.* refer to sources in the 14-h field.

\begin{figure*}
 \subfigure[\label{fig1a}]{\psfig{file=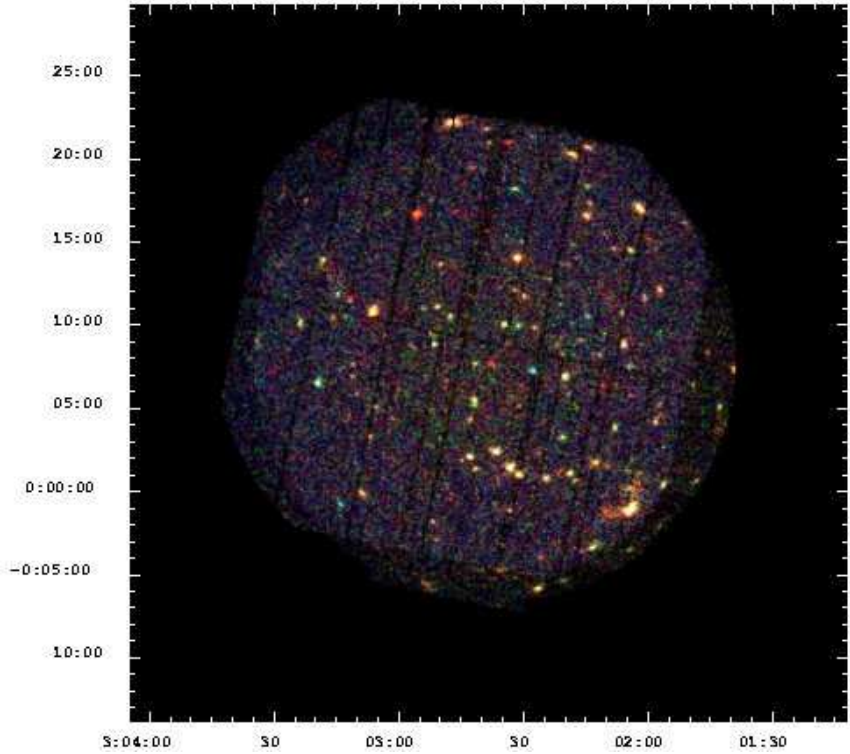,width=11.3cm,height=10cm}}
 \subfigure[\label{fig1b}]{\psfig{file=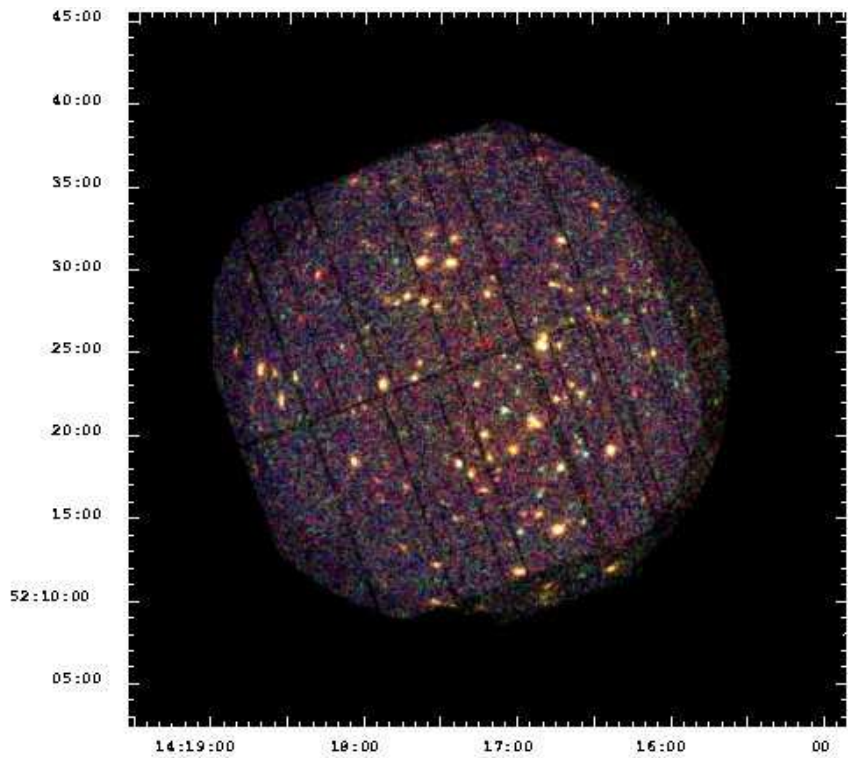,width=11.3cm,height=10cm}}
 \caption{\label{fig1} False colour X-ray image of the 3-h
 (\ref{fig1a}) and 14-h (\ref{fig1b}) fields.  Each field has one
 exposure but the images show data from all three X-ray cameras on
 {\em XMM\/}.  Soft X-rays are red ($0.5-1.5~keV$), medium are green
 ($1.5-3.5~keV$) and hard are blue ($3.5-10~keV$).}
\end{figure*}

Figure~\ref{fig1} shows false colour images of the two fields
considered in this work.  Lowest energy X-rays are coloured red with
progressively higher energy X-rays being coloured green and then blue.
Sources with hard spectra therefore show up blue in these images and
soft sources appear red.  All the extended sources detected are in the
3-h field and the majority are concentrated in the diffuse red patch
visible in the lower right hand corner of 3-h image, surrounding a
bright QSO (source 3.1 in table~\ref{3-hXray} \&~\ref{3-hIDs}).  This
could be indicative of a galaxy cluster and if the QSO is part of the
cluster then the cluster has a redshift of 0.641.  Unfortunately
because the QSO is so bright it is hard to tell if it actually lies
within a cluster, or whether the diffuse emission is simply an effect
due to the broadening of the {\em XMM\/} point spread function towards
the edge of the map.  It is also unfortunate that this particular
source lies off the edge of the deep optical map we use to identify
the X-ray sources (see below), and so an optical cluster search of
this region is not possible at this time.  Digitized Sky Survey images
of this region do not show any evidence for a galaxy cluster but do
show the optical counterpart for the QSO.

\begin{figure*}
 \subfigure[\label{countsa}]{\psfig{file=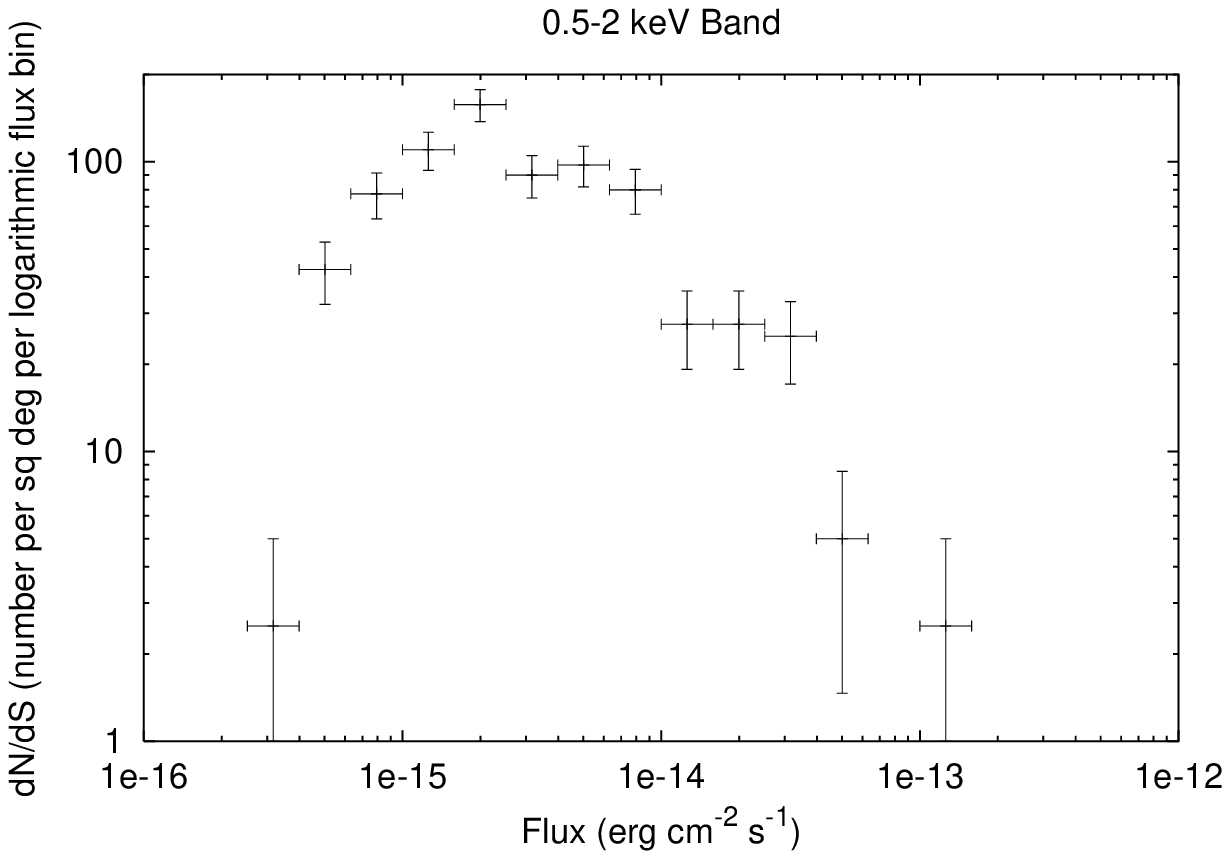,width=8.4cm,height=5.8cm}}
 \subfigure[\label{countsb}]{\psfig{file=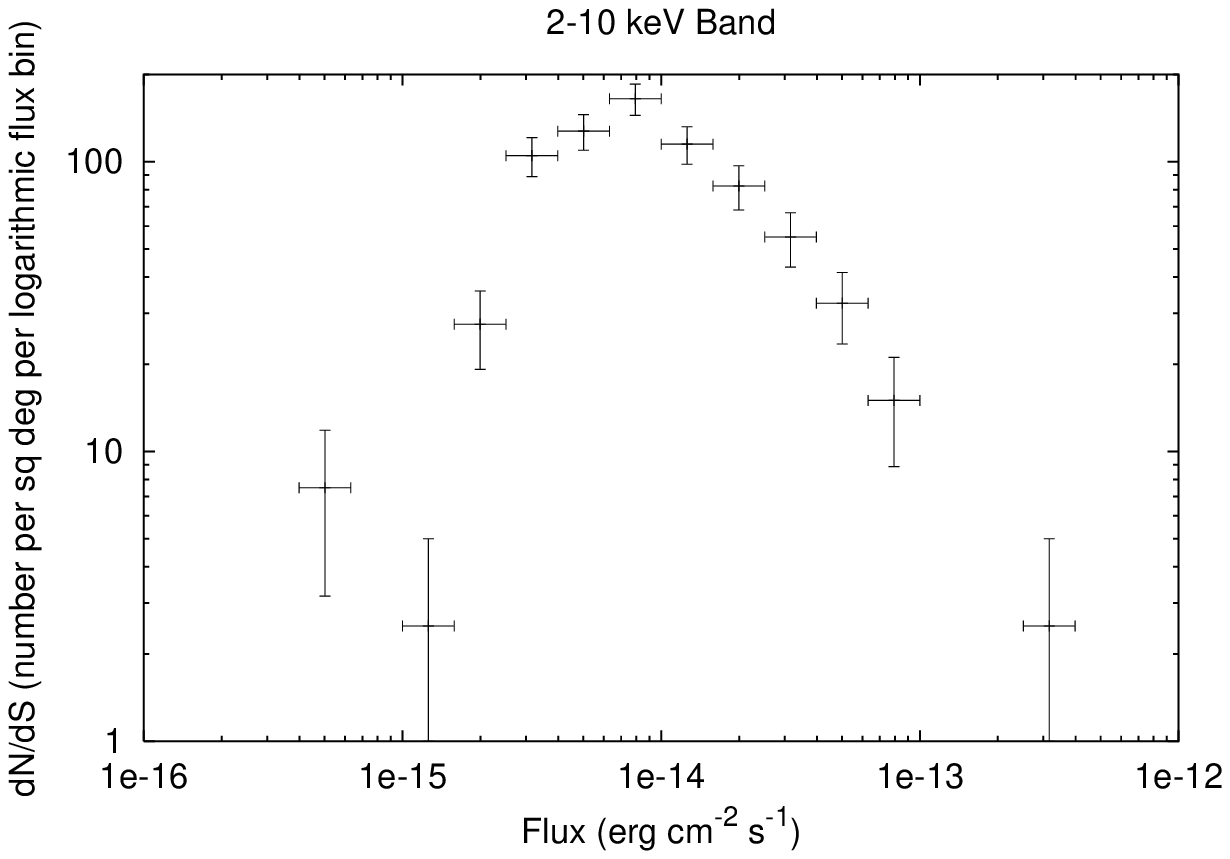,width=8.4cm,height=5.8cm}}
 \caption{\label{counts}Differential source counts for the combined
 3-h and 14-h field X-ray sources.  Incompleteness causes the source
 counts to turn over at $\sim 1.5$ and $\sim 6 \times
 10^{-15}~erg~cm^{-2}~s^{-1}$ in the soft and hard bands respectively.
 Source fluxes are calculated from the source detection procedure
 outlined in the text.  Horizontal bars indicate the logarithmic flux
 ranges over which the sources are binned.  Vertical error bars are
 the square root of the number of sources in each bin.}
\end{figure*}

Figure~\ref{counts} shows the differential source counts versus both
soft and hard band flux, for all the sources detected in the two
fields.  These plots clearly demonstrate the effect of incompleteness
at lower fluxes where the source counts drop off dramatically.  This
effect begins to become important at fluxes of 1.5 and $6 \times
10^{-15}~erg~cm^{-2}~s^{-1}$ for the soft and hard band sources
respectively.  Above these fluxes we are effectively $100~per~cent$
complete.  This is comparable in depth to, for example, the HELLAS2XMM
survey (Baldi et al. 2002), the early {\em XMM\/} Lockman Hole
observations (Hasinger et al. 2001) and serendipitous {\em Chandra\/}
observations (e.g. Gandhi et al. 2003), while reaching slightly deeper
than the Serendipitous {\em XMM\/} Survey in the AXIS field
(e.g. Barcons et al. 2002).

\section{Optical Identifications}
\subsection{XMM}
\label{xmm}
After correcting the {\em XMM\/} astrometry against known bright QSOs
the process of identifying the X-ray sources with optical counterparts
can be carried out.  This process is important for the procedures in
the following sections, especially in obtaining the redshift
distribution of the AGN.

Both survey fields are coincident with the Canada-France Redshift
Survey (CFRS) (Hammer et al. 1995; Lilly et al. 1995b) and the
Canada-France Deep Fields survey (CFDF) (McCracken et al. 2001).  The
former covers a $10\arcmin \times 10\arcmin$ section in each field
with spectroscopic redshifts for many of the galaxies, while the
latter covers almost the entirety of both and reaches 3 magnitudes
deeper ($I_{AB}(3\sigma,3\arcsec)\sim25.5$) albeit with no
spectroscopic follow-up.  We therefore use the CFDF catalogue as the
basis for our identification process and extract CFRS redshifts as
appropriate to monitor the accuracy of the photometric redshift
determination (see section~\ref{photo-z}).  The CFDF data were taken
with the {\em Canada-France-Hawaii Telescope\/} using the UH8K mosaic
camera in $B$, $V$ and $I$, with $U$ data supplied by either the CTIO
(3-h field) or the KPNO (14-h field).  Total exposure time were
typically $\sim5$ hours for $B$, $V$ and $I$, and $\sim10$ hours for
$U$.  The lengthy data reduction process is described in detail in
McCracken et al. (2001).  Of the 146(154) X-ray sources in the
3-h(14-h) fields 115(149) lie within the CFDF regions.

To determine the optical identifications of the X-ray sources we have
used the frequentist approach of Downes et al. (1986).  Since {\em
XMM\/} has a positional accuracy of $\sim2\arcsec$ (this is a worst
case situation for large off axis angles; on axis positional accuracy
is more like $1.5\arcsec$), only $1~per~cent$ of {\em XMM\/} sources
will have positions which are $>6\arcsec$ away from the object that
caused the emission.  The first step in our ID procedure was thus to
find all CFDF objects within $6\arcsec$ of the {\em XMM\/} position.
We then calculated the following statistic for each object: $$ S = 1 -
exp(-d^2 \pi n(<m)) $$ where $d$ is the offset between the {\em XMM\/}
position and that of the optical object, and $n(<m)$ is the surface
density of optical objects brighter than the magnitude ($m$) of the
possible association.  It may appear that this
statistic gives the probability that the candidate object is a
foreground or background object and is not physically related to the
{\em XMM\/}.  However, $S$ is not a probability because it doesn't
take into account galaxies that are fainter than the magnitude of the
candidate galaxy, and that {\em might\/} have had a lower value of
$S$.  Therefore this possibility needs to be taken into account when
deriving the sampling distribution of $S$. Downes et al. (1986)
describe an analytic way to do this.  The end result is a true
probability value, $P'$.  Typically a value of $P'$ is several times
higher than the equivalent $S$ value.  In all but two cases, we chose
the CFDF object with the lowest value of $P'$ as the most likely
association.  In these two exceptions, the galaxy with the lowest
value of $P'$ was close to $6\arcsec$ away from the {\em XMM\/}
position, and we preferred the candidate with a slightly higher value
of $P'$ but which was much closer to the {\em XMM\/} position (these
two IDs are confirmed by the {\em Chandra\/} X-ray positions, sources
14.15 \& 14.50).  Table~\ref{table1} gives the statistics for our
candidate identifications.
 
A consequence of this method is that because fainter objects are more
numerous, they will have higher $P'$ values than brighter objects at
the same offset.  Therefore, relatively optically faint objects are
seldom identified with X-ray sources, unless they are very close to
the X-ray position.  For example, at the optical completeness limit of
$I_{AB}=25.5$ an object at an offset of $0.8\arcsec$ will have
$P'=0.15$, which is the same $P'$ as a 20.6 magnitude object at
$6\arcsec$ offset.

\subsection{The {\em Chandra\/} Training Set}
\label{chandra}
We initially chose a $P'$ value of 0.1 as being our dividing line
between identifications and objects that are likely to be physically
unrelated to the X-ray source.  The number of spurious identifications
can be estimated by simply adding up the values of $P'$ for objects
with $P'<0.1$.  This is $\sim 2$ in the 3-h field and $\sim 3$ in the
14-h field.  In the two fields, 181 sources have $P'<0.1$, which is
$68~per~cent$ of the {\em XMM\/} sources for which there are deep CFDF
images.  The error rate of false associations is
$5/181\sim3~per~cent$.

We were able to refine our identification criteria using the fact that
part of the 14-h field has also been surveyed with {\em Chandra\/}
(the NE quadrant).  The {\em Chandra\/} data are not the focus of this
paper but they are summarised here: The data were taken in August 2002
using the ACIS-I instrument and were reduced using the standard CIAO
v2.3 data reduction software.  The total good exposure time after
screening was $158~ks$.  Source detection was performed using the CIAO
wavdetect algorithm (Freeman et al. 2002), run on images in the $0.5-8$,
$0.5-2$, $2-8$ and $4-8~keV$ bands, using a false source probability of
$10^{-7}$.  Full details of the {\em Chandra\/} observations are 
given in Nandra et al. (2004, in preparation).

Within the {\em Chandra\/} FoV there are 63 {\em XMM\/} sources, 55 of
which were also detected by {\em Chandra\/} within $10\arcsec$ of the
{\em XMM\/} position.  We performed a similar ID process to that
employed above using these new positions, and succeeded in identifying
51 of the 55 {\em Chandra\/} sources.  Two unidentified sources were
also unidentified in the {\em XMM\/} analysis, and are essentially
blank fields with no CFDF objects lying within $6\arcsec$ of either
the {\em XMM\/} or {\em Chandra\/} position (sources 14.54 \& 73).  Of
these 51 sources, 42 had previously been identified by {\em XMM\/}.
40 were identified as the same object by both {\em XMM\/} and {\em
Chandra\/}; the remaining 2 had different IDs (sources 14.10 \& 149).
However, in one of these 2 cases the {\em Chandra\/} ID was the second
best {\em XMM\/} ID (14.149) (the {\em XMM\/} IDs are listed in
tables~\ref{14-hIDs} \&~\ref{14-hopt}).  The other 9 sources were
securely identified by {\em Chandra\/} but not by XMM, so these are
considered `new' IDs (sources 14.65, 80, 85, 90, 102, 114, 115, 122,
129)

Given the expected number of spurious {\em XMM\/} IDs for the whole
14-h field (106 identified sources) is $\sim3$ we would expect 1-2
spurious IDs in the subsample covered by the {\em Chandra\/} FoV.  We
found 2 IDs that were wrong in this sample and so feel confident that
our estimate of $\sim3$ spurious {\em XMM\/} IDs in this field is
accurate.

We relaxed the selection criteria for the {\em XMM\/} ID candidates to
see if we could find more identifications for the {\em XMM\/} sources
without significantly increasing the number of false associations.  By
increasing the cut-off to $P'<0.15$ a further 5 {\em XMM\/} sources
within the {\em Chandra\/} FoV are identified.  Four of these are
judged to be correct (14.85, 90, 102 \& 114) given the {\em Chandra\/}
ID and one is incorrect (14.115).  Extrapolating to our entire survey,
we estimate that by relaxing our $P'$ criterion we gain 22 additional
identifications, of which probably $\sim5$ are inaccurate. For the
rest of this work IDs with $P'<0.15$ are considered secure.

To summarise: with this new $P'$ threshold we identify 84 out of 115
sources in the 3-h field and 119 out of 149 sources in the 14-h field.
One extra QSO lies outside the 3-h CFDF map but is coincident almost
exactly with an {\em XMM\/} source and so is identified as such.  An
additional QSO lies on a chip boundary in the 3-h field and is assumed
to be responsible for the X-ray emission detected to either side of
the boundary (sources 3.7 and 19, see table~\ref{3-hXray} and the very
top of figure~\ref{fig1a}, hereafter referred to as source 7), so in
total 86 3-h sources are identified.  In the 14-h field the {\em
Chandra\/} positions succeeded in identifying an extra 4 sources
(14.65, 80, 122 \& 129), bringing the total number of identified
sources in this field to 123.  Out of the {\em XMM\/} sources within
the area of the CFDF, we have identified $75~per~cent$ of the sources
in the 3-h field and $83~per~cent$ of the sources in the 14-h field.
Only a small part of the difference between the two fields are the
{\em Chandra\/} positions that exist for some of the 14-h {\em XMM\/}
sources.  We expect of our 209 identifications, 10 are incorrect.

There are 4 new {\em Chandra\/} IDs and 3 IDs that were changed when
{\em Chandra\/} positions were used rather than {\em XMM\/} positions
(after increasing the $P'$ limit to 0.15).  These 7 sources, that were
not possible to identify using {\em XMM\/} positional data but which
were possible to identify using {\em Chandra\/} positions (sources
14.10, 65, 80, 115, 122, 129 \& 149), give us an insight into the
properties of the remaining 58 unidentified {\em XMM\/} sources.  The
X-ray fluxes of the unidentified XMM sources cover a large range of
fluxes (see figure~\ref{ratio}), but the median $I$ magnitude of the
new IDs is 23.6, cf. the median $I$ magnitude for the other {\em
XMM\/} IDs is 21.2 (range: 11.5 to 25.5), nearly 10 times brighter.
This is a consequence of the effect described at the end of
section~\ref{xmm}.  Section~\ref{photo-z} describes the redshift
information obtained for the IDs and it appears that these {\em
Chandra \/} IDs lie, in general, at higher redshifts than most of the
{\em XMM\/} IDs, which would partially explain their relative optical
faintness.

\section{X-ray to Optical Flux Ratios}
\label{flux}

\begin{figure*}
 \subfigure[\label{ratioa}]{\psfig{file=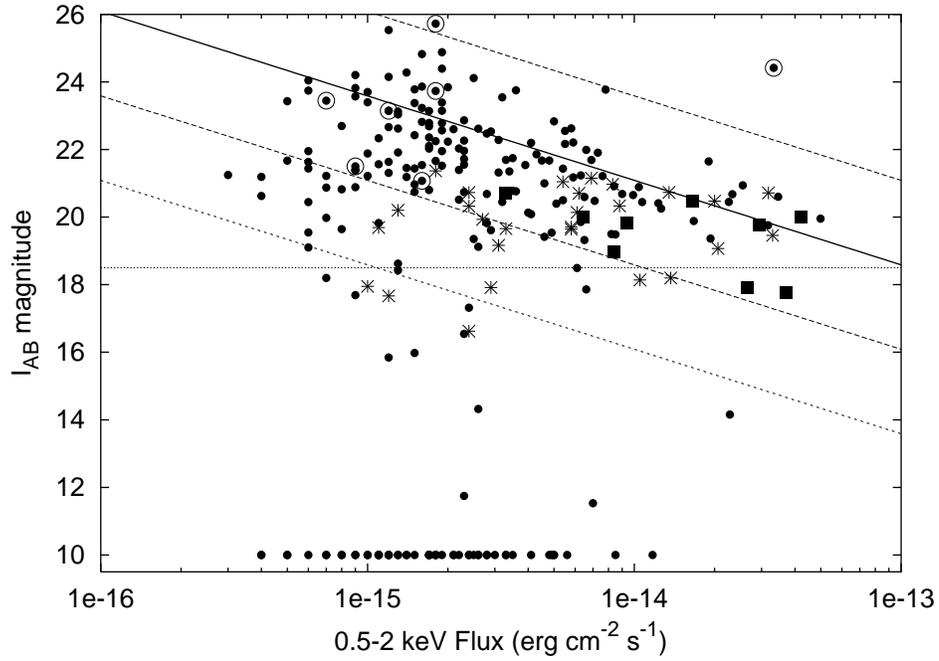,width=12.6cm,height=8.9cm}}
 \subfigure[\label{ratiob}]{\psfig{file=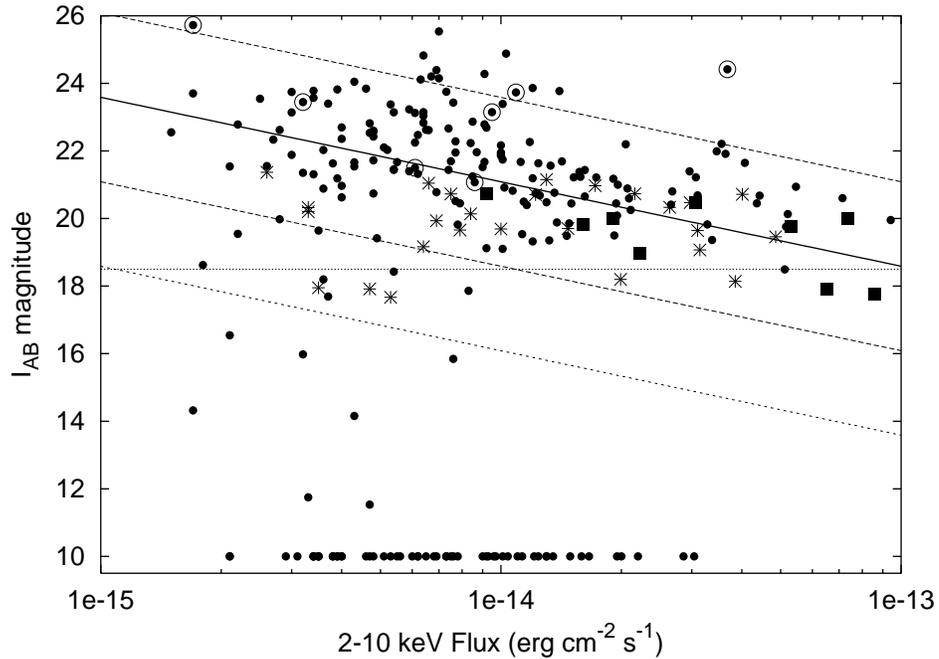,width=12.6cm,height=8.9cm}}
 \caption{\label{ratio}X-ray to optical flux ratio for all sources
 detected in both the 3 and 14-h fields.  Lines of constant flux ratio
 are plotted; solid line - $log(f_{X}/f_{I})=0$; longer dashed lines
 $\pm1$ and shorter dashed line $-2$.  AGN tend to occupy the region
 between the dashed lines, quiescent galaxies lie mostly below the
 dotted line while a mixture, including starburst galaxies, occupy the
 region in between.  The dotted line at I$_{AB}=18.5$ shows the
 saturation limit of the CFDF and so magnitudes brighter than this are
 likely to be underestimated.  Solid squares - known QSOs; asterisks -
 identifications with a stellar profile.  Unidentified sources, within
 the optical coverage, are placed at I$_{AB}=10$.  The sources
 identified using {\em Chandra\/} positions, including the 3 with
 alternative {\em Chandra\/} IDs, are ringed with larger circles
 (sources 14.10, 65, 80, 115, 122, 129 \& 149, see end of
 tables~\ref{14-hIDs} \&~\ref{14-hopt}).}
\end{figure*}

A convenient way of discriminating between different classes of X-ray
source is the ratio between their X-ray and optical flux.
Figure~\ref{ratio} shows the total $I$ band magnitude (measured using
a variable aperture to encompass the total flux of each object) versus
the X-ray flux for all the identified sources in the 3 \& 14-h fields.
The X-ray flux is calculated assuming a photon index $\Gamma = 1.7$.
The $I$ band magnitude is related to the flux in this band, $f_{I}$,
by $log~f_{I} = -0.4 I_{AB} - 5.57$, where $f_{I}$ has the units
$erg~cm^{-2}~s^{-1}$.  Lines of constant X-ray to optical flux are
plotted for comparison.  AGN tend to occupy the space between the
$log(f_{X}/f_{I}) = \pm 1$ lines while quiescent galaxies mostly lie
below the $log(f_{X}/f_{I}) = - 2$ line, with a mixture in between.

Barger et al. (2002,2003) have plotted similar diagrams for the {\em
Chandra\/}-Deep Field North survey, an X-ray sample approximately ten
times fainter than our own.  In the {\em Chandra\/} survey the median
optical apparent magnitude of X-ray sources flattens off at low X-ray
fluxes, bringing the majority of sources below the AGN region on the
$log(f_{X}/f_{I})$ plot.  However, at the flux limit of our survey we
are still predominantly detecting AGN with only a minor contribution
from quiescent galaxies.  Additionally, the redshift distribution of
our identified sources (see section~\ref{photo-z}) places the majority
of the AGN in our survey at $z<1$ which is the period of peak
formation of super-massive blackholes with low accretion rates (Cowie
et al. 2003).  These two facts mean that medium-deep surveys such as
ours are well placed to study this important period of growth for
intermediate luminosity AGN, without the need for very deep surveys,
which are able to probe much earlier times in the evolution of AGN and
study the X-ray properties of more `normal' galaxies.

In figure~\ref{ratio}, the extra sources identified by {\em
Chandra\/}, but not by {\em XMM\/}, in the 14-hr field all reside in
the higher $log(f_{X}/f_{I})$ regions.  This suggests that they are
AGN rather than starbursts or quiescent galaxies.  Given that the {\em
XMM\/} unidentified sources are in general optically fainter than the
identified ones ($I_{AB}>22$, see section~\ref{chandra} and end of
section~\ref{xmm}), and that their X-ray fluxes are similar, this
implies that the unidentified X-ray sources are most likely AGN too,
with high $f_{X}/f_{I}$ ratios.  One interesting point to note is that
source 14.10 has a different {\em Chandra\/} ID to the one given by
the {\em XMM\/} position; it is the {\em Chandra\/} ID that is
plotted in these figures.  However, the {\em Chandra\/} ID is
significantly fainter than the {\em XMM\/} ID ($I_{AB}=24.4$ cf. 19.0)
and so this source now has an extreme $log(f_{X}/f_{I})$ value of
$\sim2$ (cf. $\sim-0.3$ for the {\em XMM\/} ID).  We assume in this
paper that the {\em Chandra\/} ID is the correct one but given this
extreme flux ratio it is possible that {\em XMM\/} has correctly
identified this source, rather than {\em Chandra\/}.

In addition to the known QSOs in these fields, 27 of the
identifications have stellar optical profiles.  Figure~\ref{ratio}
shows that most of these lie in the AGN part of the diagram,
suggesting that they are QSOs rather than stars.

\section{Photometric Redshifts}
\label{photo-z}
Only a handful of the CFDF IDs have spectroscopic redshifts.  Including
known QSOs outside the CFRS regions there are 13(6) X-ray sources with
spectroscopic redshifts in the 3-h(14-h) fields.  The vast majority of
the non-broad-line AGN do not have spectroscopic redshifts and so we
turn to photometric techniques to estimate redshifts for these.

The optical spectra of broad line AGN (QSOs) are contaminated by light
from the central engine, and so obtaining photometric redshifts for
them is problematic.  However, Gonzalez \& Maccarone (2002) have shown
that for the majority of X-ray sources, which are non-broad line AGN,
the optical spectrum is not significantly contaminated and so
photometric techniques work just as well as they do with `normal'
galaxies.  As long as the QSOs can be identified they shouldn't affect
the rest of the sample.  We therefore only use the estimated
redshifts for the identifications which do {\em not\/} have a stellar
profile.  We use two photometric redshift estimation codes in this
work, a Bayesian template fitting code called BPZ (Ben\'{i}tez 2000)
and a code developed specifically for the CFDF (Brodwin et al. 2003),
calibrated against CFRS spectroscopic data.  See the appendix for the
details and a comparison of the two codes.

\begin{figure}
\psfig{file=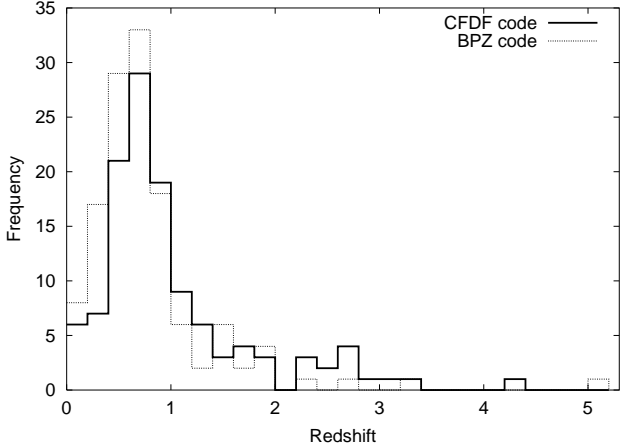,width=8.7cm,height=6.1cm}
\caption{\label{zdist}Redshift distribution of the identified X-ray
sources as measured by the two different photo-z codes.  Where a
spectroscopic redshift exists it is used in preference to the
photometric one in both histograms.  All unreliable redshifts are
excluded (i.e. those indicated with * or s in tables~\ref{3-hIDs}
\&~\ref{14-hIDs}, and saturated objects with $I_{AB}<18.5$).  The
overall shape of this distribution remains unchanged if the less
reliable redshift estimates are also included; only the normalisation
increases.}
\end{figure}

The photometry for all the IDs is listed in tables~\ref{3-hopt}
\&~\ref{14-hopt} and the results for both codes are shown in
tables~\ref{3-hIDs} \&~\ref{14-hIDs}.  Figure~\ref{zdist} shows the
redshift distribution, as measured by each code, of all the reliable
IDs that also have reliable redshift estimates, with a bin size of
$\Delta z=0.2$.  Reliable {\em photometric\/} redshifts are defined
here as unsaturated objects that have $95~per~cent$ ($\sim2\sigma$)
redshift confidence limits $<0.4(1+z)$ (CFDF code) or $P_{\Delta
z}>0.9$ (BPZ code), otherwise spectroscopic redshifts are used where
they exist; in total 129(120) estimates are reliable for the BPZ(CFDF)
code.  Despite the differences between the distributions measured by
the two different codes the overall shape of the distribution is
clear, with a peak at around $z=0.7$.  In both distributions nearly
$60~per~cent$ of the objects lie in the range $0.4\leq z < 1$.  The
median redshifts are significantly different however: 0.62 for BPZ and
0.79 for the CFDF code.  For the rest of this work the CFDF code is
assumed to be more accurate (see appendix) and so all further quoted
photometric redshifts are those given by this code.

An interesting point to note here is that the extra sources identified
by {\em Chandra\/} and not by {\em XMM\/} (see end of
table~\ref{14-hIDs}, sources 14.10, 65, 80, 115, 122, 129 \& 149) lie,
in general, at higher redshifts than the majority of the {\em XMM\/}
identified sources.  If all the unidentified sources lie at higher
redshifts than all of the other sources, then the median redshift of
the total increases to $z\sim1.1$.

\section{Absolute Magnitudes - Galaxy Types}
\begin{figure}
\psfig{file=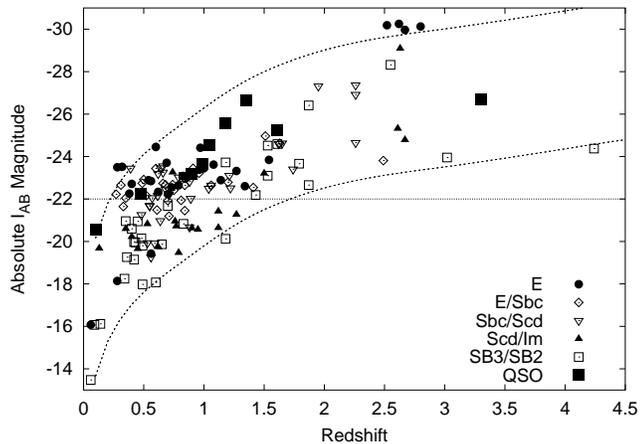,width=8.7cm,height=6.1cm}
\caption{\label{absmag}Absolute $I_{AB}$ magnitude vs. redshift for
the identified sources (148 sources after removal of stellar and
saturated objects).  The horizontal line is the approximate position
of $M^{*}_{I}$.  The first four galaxy types are taken from Coleman,
Wu \& Weedman (1980), although some interpolation is used to create
intermediate templates, and the starburst symbol represents both the
SB3 and SB2 types from Kinney et al. (1996).  QSOs have been plotted
using the best fit template for the K correction, in general the
bluest starburst.  Upper and lower curves are the approximate
saturation limit and completeness limit of the optical data
($I_{AB}=18$ and 24.5 respectively), calculated for an Scd galaxy (due
to larger K corrections some ellipticals lie above the bright limit
for Scd galaxies).}
\end{figure}

Figure~\ref{absmag} shows the absolute $I_{AB}$ magnitude plotted
against redshift.  The different symbols represent the best fitting
template determined for each galaxy by the CFDF code, using six band
photometry.  Although the code uses 15 templates for greater accuracy
each symbol here represents a small range of templates for clarity.
In general the two photo-z codes agree reasonably well as to the best
fitting galaxy type.

\begin{figure}
\psfig{file=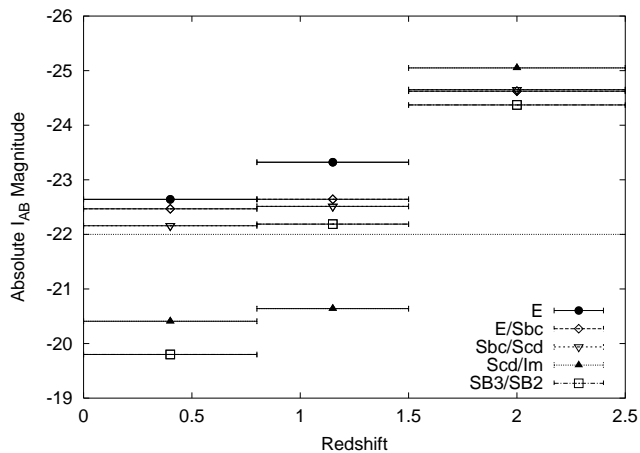,width=8.7cm,height=6.1cm}
\caption{\label{absmagmed}Median values for the galaxy types in
figure~\ref{absmag}.  The highest z bin includes all sources with
z~$>1.5$.  The group of Elliptical galaxies at M~$\sim -30$ in
figure~\ref{absmag} is off the vertical scale in this plot.  The
horizontal line is the approximate position of $M^{*}_{I}$.}
\end{figure}

The X-ray sources all lie in a band defined at the faint limit by the
limiting magnitude of the optical data, and at the bright limit by the
saturation magnitude.  Objects brighter than this magnitude do not
have reliable photometric redshifts and so do not appear in this plot.
Aside from these selection effects there are several other trends
apparent here.  Apart from QSOs, in general, at lower redshifts, the
bluer galaxy types occupy the region near the faint limit while
progressively redder galaxies occur at brighter magnitudes (see
figure~\ref{absmagmed}).  However, this trend breaks down at higher
redshift where there are fewer sources, and errors in the photometry
are likely to be more important.  There is no clear domination of one
galaxy type over any other, indicating that AGN have no preference
when it comes to the morphology of their host galaxies.  Nor is there
any apparent preference for optical luminosity of the host galaxy,
unlike the narrow absolute magnitude range preferred by the starburst
galaxies detected in $\mu Jy$ radio surveys (Chapman et al. 2003).
X-ray sources occupy the whole optical luminosity range available to
them in this plot.  There are four apparently very luminous
ellipticals at $z>2.5$ (sources 3.32, 3.90, 3.92 \& 14.31) which may
be erroneous identifications.  The CFDF code becomes less reliable
above a redshift of 1.3 (Brodwin et al. 2003) and so it is possible
that these sources actually lie at lower redshifts (in fact the BPZ
code places three of these sources at $z<1$, see tables~\ref{3-hIDs}
\&~\ref{14-hIDs}, and classifies them as spirals; likely a consequence
of this code using a magnitude based prior) and so are consequently of
less extreme luminosity.  With this in mind, high redshift sources
should be viewed with some caution.

The extra sources identified by {\em Chandra\/} but not by {\em XMM\/}
in the 14-hr field also cover a wide range in galaxy types.  The two
higher redshift sources are the bluest galaxy types while the two
lowest redshift sources are the reddest types.  Three of the four hug
the lower luminosity limit, a consequence of their relative optical
faintness.

\begin{figure}
\psfig{file=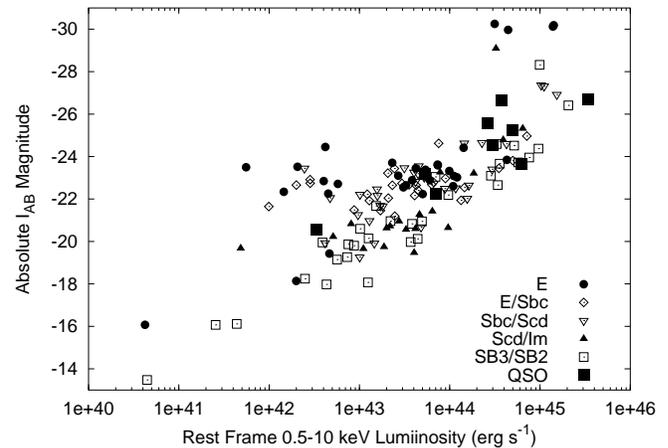,width=8.7cm,height=6.1cm}
\caption{\label{absmagvslum}Absolute $I_{AB}$ magnitude vs. X-ray
luminosity calculated from the $0.5-10~keV$ flux for the same sources
as figure~\ref{absmag}.  The X-ray luminosity is K-corrected assuming
an intrinsic power law slope with photon index $\Gamma=1.7$.}
\end{figure}

Converting the $0.5-10~keV$ X-ray flux of the identified sources into
X-ray luminosity gives us figure~\ref{absmagvslum}.  Although the
striking correlation here is possibly dominated by the same selection
effects seen in figure~\ref{absmag}, it is rather reminiscent of the
Magorrian relation (Magorrian et al. 1998), with black-hole mass
represented by X-ray luminosity and bulge mass represented by optical
luminosity.  Whether this correlation is real or not depends on
exactly where the optically faint and saturated sources lie in this
plot.  We would expect optically faint sources to fall in the lower
right part of this plot and the saturated sources to fall in the upper
left part, effectively smearing out the correlation.  However, if the
optically faint sources are {\em not\/} at much higher redshifts than
the identified sources (contrary to our arguments above) then both
their X-ray and optical luminosities will be low, placing them amongst
the sources plotted here.  The very luminous ellipticals, mentioned
above, also appear in this plot, slightly above the general trend,
again suggesting that they have been misclassified (as have,
potentially, a group of lower luminosity ellipticals, also lying away
from the trend).

\begin{figure}
\psfig{file=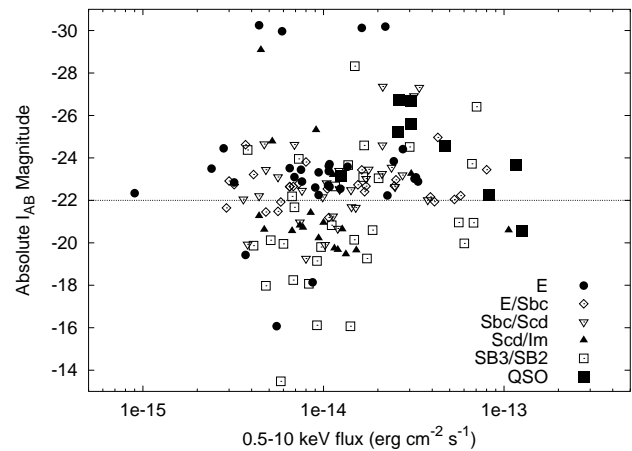,width=8.7cm,height=6.1cm}
\caption{\label{fluxvsabsmag}Absolute $I_{AB}$ magnitude vs. total
X-ray flux for the same sources as figure~\ref{absmag}.  The
horizontal line is the approximate position of $M^{*}_{I}$.}
\end{figure}

Plotting these sources in a different way illustrates what type of
objects contribute to the XRB.  For this discussion we assume the XRB
to have a spectrum of
$I(E)=11E^{-0.4}~keV~s^{-1}~cm^{-2}~sr^{-1}~keV^{-1}$ (McCammon \&
Sanders 1990; Fabian \& Barcons 1992), although the overall
normalisation is still somewhat uncertain.  Figure~\ref{fluxvsabsmag}
shows absolute $I_{AB}$ magnitude vs. X-ray flux with the same symbols
as in figure~\ref{absmag}.  The 300 sources in this survey (assuming
the majority of the `stars' are misidentified QSOs) contribute $\sim
51~per~cent$ to the XRB in the $0.5-10~keV$ range, while the 148
sources included in these figures (ie. the securely identified sources
with redshift estimates) provide $\sim 27~per~cent$.  Of this
$27~per~cent$, sources brighter than $M^{*}_{I}$ contribute
$69~per~cent$ while fainter sources contribute $31~per~cent$.  This
calculation shows that the XRB is not dominated by the most optically
luminous galaxies, but that a significant contribution comes from
galaxies with fairly low optical luminosity.

\section{Concluding Remarks}
We have presented source catalogues for a survey, using the {\em
XMM-Newton\/} X-ray telescope, of $\sim0.4$ square degrees of sky.  We
show that reliable identifications can be obtained for
$\sim75~per~cent$ of the {\em XMM\/} sources using {\em XMM\/}
positions alone.  Those sources that cannot be identified using {\em
XMM\/} positions alone are optically fainter ($I_{AB}>22$) than most
of the identified ones, and are likely to be AGN at generally higher
redshifts.  We have obtained the following results:

\begin{itemize}

\item The flux ratio $f_{X}/f_{opt}$ of the sources in our survey show
that they are predominantly AGN.

\item The optical properties of the AGN span a large range of absolute 
magnitudes, in contrast to the result found for the starburst galaxies
detected in $\mu Jy$ radio surveys, which tend to have a very narrow
range of absolute magnitudes (Chapman et al. 2003).

\item AGN are found in host galaxies spanning the full range of Hubble
types, with no clear preference.

\item For the identified X-ray sources with good redshifts there is a
strong correlation between optical and X-ray luminosity, reminiscent
of the Magorrian relation between black-hole mass and bulge mass.
However, this may be due to selection effects.

\item The redshift distribution of the AGN shows a clear peak at
$z\sim0.7$.

\end{itemize}

The last result supports other recent studies (Barger et al. 2003)
that show the peak formation of super-massive black holes occurred at
relatively recent times ($z<1$).  Medium-deep X-ray surveys such as
ours, which resolve a large fraction of the XRB but are still
dominated by AGN, are able to probe this epoch effectively.

We will use the results from this paper to calculate the X-ray
luminosity function, and determine its evolution with redshift, in the 
next paper in this series.

\section{Catalogue}

The following tables are a sample of the full catalogue, which can be
found in the electronic version of this paper.  It is split into 3
sections for each of the two fields in this survey.  The first tables
for each field (\ref{3-hXray} \&~\ref{14-hXray}) contain the positions
and fluxes of the X-ray sources as measured by the source detection
software.  The second tables (\ref{3-hIDs} \&~\ref{14-hIDs}) have the
identification information for all of the good ID candidates
($P'<0.15$) including the CFDF catalogue number, the ID position, the
distance between the ID and its corresponding X-ray source, $P'$ value
and redshift information.  The final tables (\ref{3-hopt}
\&~\ref{14-hopt}) show the photometry for each good ID.  The X-ray
sources are ordered by total number of counts detected in the full
X-ray band, greatest first.  Due to vignetting this order is
approximately but not exactly the same as the flux order.  Source 23
in the 14-h field is detected by SCUBA at $850\mu$m and is discussed
in Paper-I in more detail.

All six tables can also be found in their entirety, in electronic text
format, at the following address:
http://www.astro.cardiff.ac.uk/pub/Timothy.Waskett/

\begin{table*}
\begin{minipage}{17.5cm}
\caption{\label{3-hXray}X-ray properties of the 3-h field XMM
sources. Sources 7 and 19 are in fact the same source split into
two due to it lying on a PN chip gap (*).}
\centering
\begin{tabular}{cccccccccl} \hline
\multicolumn{1}{c}{XMM}&
\multicolumn{1}{c}{R.A.[fk5]}&
\multicolumn{1}{c}{Dec.[fk5]}&
\multicolumn{1}{c}{$0.5-2~keV^{a}$}&
\multicolumn{1}{c}{$2-10~keV^{a}$}&
\multicolumn{1}{c}{$0.5-10~keV$\footnote{Flux in units of
$10^{-15}~erg~s^{-1}~cm^{-2}$, based on a photon index of 1.7.}}&
\multicolumn{1}{c}{PN HR$^{b}$}&
\multicolumn{1}{c}{M1 HR$^{b}$}&
\multicolumn{1}{c}{M2 HR\footnote{Hardness ratio given by source
detection procedure, one for each X-ray camera.  Marked with `?' if
not detected or a bad measurement.}}&
\multicolumn{1}{l}{Notes\footnote{- = lies outside CFDF map; e =
extended source (X-ray property); q = known QSO; s = object with a
stellar profile, from $P'<0.15$ list (q and s are optical properties).}} \\ \hline
1 & 45.52820 & -0.02260 & 119.0 $\pm$ 2.6 & 262.9 $\pm$ 9.7 & 382.9
$\pm$ 10.1 & -0.4 $\pm$ 0.0 & -0.4 $\pm$ 0.0 & -0.4 $\pm$ 0.0 & - q \\
2 & 45.78054 & 0.17228 & 33.0 $\pm$ 1.2 & 48.6 $\pm$ 3.8 & 81.7 $\pm$ 4 & -0.6 $\pm$ 0.0 & -0.5 $\pm$ 0.1 & -0.5 $\pm$ 0.1 & s \\
3 & 45.64490 & 0.01902 & 23.3 $\pm$ 1.0 & 44.3 $\pm$ 3.2 & 67.7 $\pm$ 3.4 & -0.5 $\pm$ 0.0 & -0.5 $\pm$ 0.1 & -0.4 $\pm$ 0.1 &  \\
4 & 45.51813 & 0.27387 & 33.4 $\pm$ 1.5 & 76.3 $\pm$ 6.1 & 109.7 $\pm$ 6.3 & -0.5 $\pm$ 0.1 & -0.3 $\pm$ 0.1 & -0.3 $\pm$ 0.1 & - \\
5 & 45.63815 & 0.22543 & 16.5 $\pm$ 0.8 & 30.7 $\pm$ 2.7 & 47.2 $\pm$ 2.9 & -0.6 $\pm$ 0.1 & -0.2 $\pm$ 0.1 & -0.4 $\pm$ 0.1 & q \\
6 & 45.73817 & 0.26816 & 22.8 $\pm$ 1.2 & 4.3 $\pm$ 2.2 & 28.2 $\pm$ 2.6 & -1.0 $\pm$ 0.0 & -0.8 $\pm$ 0.1 & -1.0 $\pm$ 0.4 &  \\
7 & 45.70564 & 0.35812 & 37.1 $\pm$ 2.3 & 85.8 $\pm$ 10.1 & 126.3 $\pm$ 10.3 & -0.4 $\pm$ 0.1 & 1.0 $\pm$ 109.0 & ? $\pm$ ? & q* \\
8 & 45.58501 & 0.32717 & 20.6 $\pm$ 1.2 & 31.4 $\pm$ 4.2 & 53.6 $\pm$ 4.4 & -0.5 $\pm$ 0.1 & -0.6 $\pm$ 0.1 & -0.6 $\pm$ 0.1 & s \\
9 & 45.65876 & 0.03438 & 7.0 $\pm$ 0.5 & 23.0 $\pm$ 2.3 & 30.5 $\pm$ 2.4 & -0.1 $\pm$ 0.1 & -0.3 $\pm$ 0.1 & -0.3 $\pm$ 0.1 & q \\
10 & 45.59179 & 0.10849 & 6.4 $\pm$ 0.5 & 19.1 $\pm$ 2.0 & 26.1 $\pm$ 2 & -0.2 $\pm$ 0.1 & -0.5 $\pm$ 0.1 & -0.1 $\pm$ 0.1 & q \\ \hline
\end{tabular}
\end{minipage}
\end{table*}

\begin{table*}
\begin{minipage}{17.5cm}
\caption{\label{14-hXray}X-ray properties of the 14-h field XMM
sources.}
\centering
\begin{tabular}{cccccccccl} \hline
\multicolumn{1}{c}{XMM}&
\multicolumn{1}{c}{R.A.[fk5]}&
\multicolumn{1}{c}{Dec.[fk5]}&
\multicolumn{1}{c}{$0.5-2~keV^{a}$}&
\multicolumn{1}{c}{$2-10~keV^{a}$}&
\multicolumn{1}{c}{$0.5-10~keV$\footnote{Flux in units of
$10^{-15}~erg~s^{-1}~cm^{-2}$, based on a photon index of 1.7.}}&
\multicolumn{1}{c}{PN HR$^{b}$}&
\multicolumn{1}{c}{M1 HR$^{b}$}&
\multicolumn{1}{c}{M2 HR\footnote{Hardness ratio given by source
detection procedure, one for each X-ray camera.  Marked with `?' if
not detected or a bad measurement.}}&
\multicolumn{1}{c}{Notes\footnote{- = lies outside CFDF map; c = lies
within {\em Chandra\/} map; d = detected by {\em Chandra\/} (c and d
are X-ray properties); q = known QSO; s = object with a stellar
profile, from $P'<0.15$ list (q and s are optical properties).}} \\ \hline
1 & 214.2072 & 52.42472 & 34.6 $\pm$ 0.9 & 71.4 $\pm$ 3.1 & 106.0 $\pm$ 3.2 & -0.5 $\pm$ 0.0 & -0.4 $\pm$ 0.0 & -0.4 $\pm$ 0.0 & c d \\
2 & 214.4009 & 52.50781 & 42.2 $\pm$ 1.2 & 73.4 $\pm$ 4.0 & 115.9 $\pm$ 4.1 & -0.5 $\pm$ 0.0 & -0.5 $\pm$ 0.0 & -0.5 $\pm$ 0.0 & c d q \\
3 & 214.1816 & 52.24290 & 49.9 $\pm$ 1.5 & 94.2 $\pm$ 5.3 & 144.3 $\pm$ 5.5 & -0.5 $\pm$ 0.0 & -0.4 $\pm$ 0.1 & -0.4 $\pm$ 0.1 &  \\
4 & 214.3536 & 52.50655 & 29.3 $\pm$ 1.0 & 53.2 $\pm$ 3.5 & 82.6 $\pm$ 3.7 & -0.5 $\pm$ 0.0 & -0.4 $\pm$ 0.1 & -0.5 $\pm$ 0.1 & c d q \\
5 & 214.0966 & 52.32077 & 25.5 $\pm$ 1.0 & 54.6 $\pm$ 3.6 & 80.1 $\pm$ 3.8 & -0.4 $\pm$ 0.0 & -0.3 $\pm$ 0.1 & -0.4 $\pm$ 0.1 &  \\
6 & 214.4645 & 52.38579 & 19.3 $\pm$ 0.8 & 33.7 $\pm$ 2.7 & 53.1 $\pm$ 2.8 & -0.5 $\pm$ 0.0 & -0.4 $\pm$ 0.1 & -0.5 $\pm$ 0.1 & c d \\
7 & 214.2442 & 52.20099 & 31.7 $\pm$ 1.3 & 51.6 $\pm$ 4.8 & 83.8 $\pm$ 5.0 & -0.6 $\pm$ 0.0 & -0.4 $\pm$ 0.1 & -0.5 $\pm$ 0.1 &  \\
8 & 214.2543 & 52.32128 & 16.7 $\pm$ 0.7 & 13.8 $\pm$ 1.7 & 30.6 $\pm$ 1.8 & -0.8 $\pm$ 0.0 & -0.6 $\pm$ 0.1 & -0.7 $\pm$ 0.1 & c d \\
9 & 214.2152 & 52.34575 & 12.6 $\pm$ 0.6 & 21.1 $\pm$ 2.0 & 33.9 $\pm$ 2.1 & -0.6 $\pm$ 0.1 & -0.5 $\pm$ 0.1 & -0.4 $\pm$ 0.1 & c d \\
10 & 214.6612 & 52.39937 & 33.3 $\pm$ 1.5 & 36.8 $\pm$ 5.0 & 70.4
$\pm$ 5.2 & -0.7 $\pm$ 0.1 & -0.6 $\pm$ 0.1 & -0.6 $\pm$ 0.1 & c d \\ \hline
\end{tabular}
\end{minipage}
\end{table*}

\begin{table*}
\begin{minipage}{17.5cm}
\caption{\label{3-hIDs}ID properties of the 3-h field {\em XMM\/} source IDs
($P'<0.15$).  Coordinates are for the CFDF objects {\em not\/} the XMM
sources; the offset is between the {\em XMM\/} source and the CFDF
object; The first two photo-z columns ($UBVI$ \& $UBVIK$) are BPZ
estimates, * indicates that $P_{\Delta z}<0.9$ and so may be less
reliable (Ben\'{i}tez 2000); $UBVRIZ$ is the photo-z estimate given by
the CFDF code and here * also indicates a less reliable estimate
because of multiple likelihood peaks or broad errors.  Note, not all
IDs are included in the recent $UBVRIZ$ catalogue, so these sources do
not have redshift estimates in this column.  Photo-z estimates are
always unreliable for QSOs (and potentially misidentified stars) and
saturated objects ($I_{Tot}<18.5$), regardless of any other
reliability measure.  Notes have the same meaning as in
table~\ref{3-hXray}.}
\centering
\begin{tabular}{cccccccllll} \hline
\multicolumn{1}{c}{XMM}&
\multicolumn{1}{c}{CFDF}&
\multicolumn{1}{c}{R.A.[fk5]}&
\multicolumn{1}{c}{Dec.[fk5]}&
\multicolumn{1}{c}{Offset(\arcsec)}&
\multicolumn{1}{c}{$P'$}&
\multicolumn{1}{c}{$z_{sp}$}&
\multicolumn{1}{l}{$UBVI$}&
\multicolumn{1}{l}{$UBVIK$}&
\multicolumn{1}{l}{$UBVRIZ$}&
\multicolumn{1}{l}{Notes} \\ \hline
1 &  & 45.52829 & -0.02246 &  &  & 0.641 &  &  &  & q \\
2 & 48603 & 45.78065 & 0.17224 & 0.4 & 8.63E-04 &  & 0.01 &  & 2.01 & s \\
3 & 80878 & 45.64505 & 0.01880 & 1.0 & 8.64E-03 &  & 0.40 &  & 0.45 &  \\
5 & 36830 & 45.63823 & 0.22535 & 0.4 & 1.65E-03 & 1.048 & 0.30 & 0.20 & 2.22 & q \\
6 & 27229 & 45.73788 & 0.26803 & 1.1 & 9.46E-05 &  & 0.36 &  & 3.09 * &  \\
7 & 9684 & 45.70320 & 0.35877 &  &  & 0.107 & 0.08 &  & 0.38 & q \\
8 & 15331 & 45.58520 & 0.32705 & 0.8 & 1.95E-03 &  & 0.04 &  & 0.51 & s \\
9 & 78735 & 45.65896 & 0.03432 & 0.7 & 7.20E-04 & 1.350 & 0.19 &  &  & q \\
10 & 63707 & 45.59203 & 0.10864 & 1.0 & 5.86E-03 & 3.300 & 0.27 & 0.02 & 3.27 & q \\ \hline
\end{tabular}
\end{minipage}
\end{table*}

\begin{table}
\begin{minipage}{8.4cm}
\caption{\label{table1}Summary of ID statistics for both {\em XMM\/}
fields.  $P'$ values given are for the best ID where more than one
candidate lies within the $6\arcsec$ search radius.}
\centering
\begin{tabular}{ccc} \hline
\multicolumn{1}{c}{}&
\multicolumn{1}{c}{3-h}&
\multicolumn{1}{c}{14-h}\\ \hline 
$P' < 0.05$       & 59 & 82\\
$0.05 < P' < 0.1$ & 16 & 24\\
$0.1 < P' < 0.2$  & 16 & 18\\
$0.2 < P' < 0.5$  & 13 & 16\\
$P' > 0.5$        & 12 & 8\\
Blank Field      & 2  & 1\\
Outside CFDF     & 28 & 5\\ \hline
\end{tabular}
\end{minipage}
\end{table}

\begin{table*}
\begin{minipage}{17.5cm}
\caption{\label{14-hIDs}As table~\ref{3-hIDs} but for the 14-h field.
All IDs are for {\em XMM\/} sources, except the last 7 sources, of
which 4 are the extra {\em Chandra\/} IDs and 3 are the alternative
{\em Chandra\/} IDs for sources 10, 115 \& 149 (assumed to be the
correct IDs in this work).  The two BPZ columns have been corrected for
the systematic error found in this field (see appendix).  Notes have
the same meaning as in table~\ref{14-hXray}.}
\centering
\begin{tabular}{cccccccllll} \hline
\multicolumn{1}{c}{XMM}&
\multicolumn{1}{c}{CFDF}&
\multicolumn{1}{c}{R.A.[fk5]}&
\multicolumn{1}{c}{Dec.[fk5]}&
\multicolumn{1}{c}{Offset(\arcsec)}&
\multicolumn{1}{c}{$P'$}&
\multicolumn{1}{c}{$z_{sp}$}&
\multicolumn{1}{l}{$UBVI$}&
\multicolumn{1}{l}{$UBVIK$}&
\multicolumn{1}{l}{$UBVRIZ$}&
\multicolumn{1}{l}{Notes} \\ \hline
1 & 34649 & 214.2061 & 52.42517 & 0.7 & 5.05E-03 &  & 0.33 &  & 0.35 * & c d \\
2 & 56149 & 214.3996 & 52.50816 & 0.1 & 8.83E-05 & 0.985 & 0.12 & 0.01 & 0.35 & c d q \\
3 & 32209 & 214.1803 & 52.24317 & 0.4 & 9.65E-04 &  & 1.11 &  &  &  \\
4 & 50800 & 214.3523 & 52.50681 & 0.4 & 8.15E-04 & 0.479 & 0.01 & 0.01 & 1.16 & c d q \\
5 & 22314 & 214.0946 & 52.32109 & 1.4 & 1.85E-02 &  & 0.79 &  & 0.91 &  \\
6 & 62713 & 214.4630 & 52.38622 & 0.4 & 7.77E-04 &  & 0.35 &  & 0.35 & c d \\
7 & 38711 & 214.2426 & 52.20126 & 0.6 & 1.81E-03 &  & 0.25 &  &  &  \\
8 & 39972 & 214.2527 & 52.32180 & 0.8 & 3.59E-03 &  & 0.60 &  & 0.74 & c d \\
9 & 35492 & 214.2133 & 52.34607 & 1.1 & 8.10E-03 &  & 0.05 &  & 1.95 & c d \\
10 & 83085 & 214.6597 & 52.39960 & 0.5 & 6.75E-04 &  & 0.01 &  & 1.87
& c d \\ \hline
\end{tabular}
\end{minipage}
\end{table*}

\begin{table*}
\begin{minipage}{17.5cm}
\caption{\label{3-hopt}Optical properties of the 3-h field {\em XMM\/} source
IDs, as extracted from the original CFDF $UBVI$ catalogues, including CFRS
$K$ photometry where available.  All magnitudes are AB and measured in
a $3\arcsec$ diameter aperture, except for the I$_{Tot}$ magnitude which
is the total magnitude measured using a variable aperture.  This total
magnitude is used as the prior in the BPZ photometric redshift code,
while the $3\arcsec$ aperture magnitudes and errors are used as the input 
catalogue.}
\centering
\begin{tabular}{ccccccccccccc} \hline
\multicolumn{1}{c}{XMM}&
\multicolumn{1}{c}{CFDF}&
\multicolumn{1}{c}{I$_{Tot}$}&
\multicolumn{1}{c}{U}&
\multicolumn{1}{c}{$\Delta$U}&
\multicolumn{1}{c}{B}&
\multicolumn{1}{c}{$\Delta$B}&
\multicolumn{1}{c}{V}&
\multicolumn{1}{c}{$\Delta$V}&
\multicolumn{1}{c}{I}&
\multicolumn{1}{c}{$\Delta$I}&
\multicolumn{1}{c}{K}&
\multicolumn{1}{c}{$\Delta$K} \\ \hline
2 & 48603 & 19.460 & 19.983 & 0.010 & 20.049 & 0.007 & 19.534 & 0.003 & 19.631 & 0.002 &  &  \\
3 & 80878 & 20.675 & 21.677 & 0.017 & 21.995 & 0.009 & 21.185 & 0.008 & 20.895 & 0.005 &  &  \\
5 & 36830 & 20.470 & 21.978 & 0.021 & 22.340 & 0.022 & 21.442 & 0.007 & 20.755 & 0.004 & 19.05 & 0.05 \\
6 & 27229 & 14.156 & 19.903 & 0.008 & 18.154 & 0.002 & 16.867 & 0.001 & 15.310 & 0.000 &  &  \\
7 & 9684 & 17.771 & 20.683 & 0.012 & 19.184 & 0.004 & 19.143 & 0.002 & 18.295 & 0.001 &  &  \\
8 & 15331 & 19.071 & 20.343 & 0.010 & 19.986 & 0.007 & 19.575 & 0.003 & 19.217 & 0.002 &  &  \\
9 & 78735 & 17.924 & 18.909 & 0.005 & 18.654 & 0.003 & 18.481 & 0.002 & 18.082 & 0.001 &  &  \\
10 & 63707 & 19.996 & 24.437 & 0.071 & 21.355 & 0.006 & 20.537 & 0.005 & 20.243 & 0.003 & 19.93 & 0.11 \\ \hline
\end{tabular}
\end{minipage}
\end{table*}

\begin{table*}
\begin{minipage}{17.5cm}
\caption{\label{14-hopt}As in table~\ref{3-hopt} but for the 14-h
field.  All are {\em XMM\/} IDs except the 7 sources at the end, of
which 4 are the extra {\em Chandra\/} IDs and 3 are the alternative
{\em Chandra\/} IDs for sources 10, 115 \& 149 (assumed to be the
correct IDs in this work).}
\centering
\begin{tabular}{ccccccccccccc} \hline
\multicolumn{1}{c}{XMM}&
\multicolumn{1}{c}{CFDF}&
\multicolumn{1}{c}{I$_{Tot}$}&
\multicolumn{1}{c}{U}&
\multicolumn{1}{c}{$\Delta$U}&
\multicolumn{1}{c}{B}&
\multicolumn{1}{c}{$\Delta$B}&
\multicolumn{1}{c}{V}&
\multicolumn{1}{c}{$\Delta$V}&
\multicolumn{1}{c}{I}&
\multicolumn{1}{c}{$\Delta$I}&
\multicolumn{1}{c}{K}&
\multicolumn{1}{c}{$\Delta$K} \\ \hline
1 & 34649 & 20.604 & 22.083 & 0.027 & 21.833 & 0.008 & 20.906 & 0.005 & 20.887 & 0.004 &  &  \\
2 & 56149 & 19.996 & 21.011 & 0.011 & 20.814 & 0.004 & 20.206 & 0.003 & 20.146 & 0.002 & 19.25 & 0.04 \\
3 & 32209 & 19.955 & 21.499 & 0.016 & 21.650 & 0.012 & 21.687 & 0.009 & 20.375 & 0.003 &  &  \\
4 & 50800 & 19.762 & 20.918 & 0.011 & 20.549 & 0.003 & 20.338 & 0.003 & 19.932 & 0.002 & 18.62 & 0.02 \\
5 & 22314 & 20.940 & 23.925 & 0.052 & 23.580 & 0.023 & 22.796 & 0.014 & 21.181 & 0.005 &  &  \\
6 & 62713 & 19.365 & 22.842 & 0.031 & 21.814 & 0.007 & 20.807 & 0.004 & 19.632 & 0.002 &  &  \\
7 & 38711 & 19.756 & 20.889 & 0.012 & 20.609 & 0.007 & 20.529 & 0.005 & 19.986 & 0.003 &  &  \\
8 & 39972 & 19.882 & 21.883 & 0.019 & 21.622 & 0.009 & 21.131 & 0.006 & 20.266 & 0.003 &  &  \\
9 & 35492 & 20.252 & 21.836 & 0.019 & 21.512 & 0.008 & 21.141 & 0.006 & 20.690 & 0.004 &  &  \\
10 & 83085 & 19.029 & 19.341 & 0.006 & 19.402 & 0.003 & 19.050 & 0.002 & 19.163 & 0.002 &  &  \\ \hline
\end{tabular}
\end{minipage}
\end{table*}

\section*{acknowledgements}
TJW acknowledges the support of a departmental postgraduate grant.
SAE thanks the Leverhume Trust for a research fellowship.  We thank
the referee for being so thorough, which greatly clarify the paper and
also helped us to better organise our thoughts for further work.  This
paper was based on observations obtained with {\em XMM-Newton\/}, an
ESA science mission with instruments and contributions directly funded
by ESA Member States and NASA.

\section*{Appendix}
The work in this paper depends heavily on the reliability of the
photometric redshift estimation codes we use.  There are two codes
whose results are presented in this work, one by Ben\'{i}tez (2000)
which uses a Bayesian approach and template fitting technique, called
BPZ; and another that is developed by one of us specifically for the
CFDF (Brodwin et al. 2003) utilising the CFRS (Hammer et al. 1995;
Lilly et al. 1995b) to calibrate the template fitting.  These two
codes are slightly different and each have their strengths.  This
appendix is concerned with the reliability testing of these two codes.
For a more detailed discussion of the CFDF code, and its reliability
when compared against the CFRS spectroscopic sample, refer to Brodwin
et al. (2003).

For reasons of timing the input to BPZ is from an older version of the
CFDF catalogues than that employed for the specific CFDF code.
Therefore this should be taken into account when comparing the two
codes.

\subsection*{BPZ Photometric Redshift Estimation Code}
\label{BPZ}

\begin{figure*}
 \subfigure[\label{photspeca}]{\psfig{file=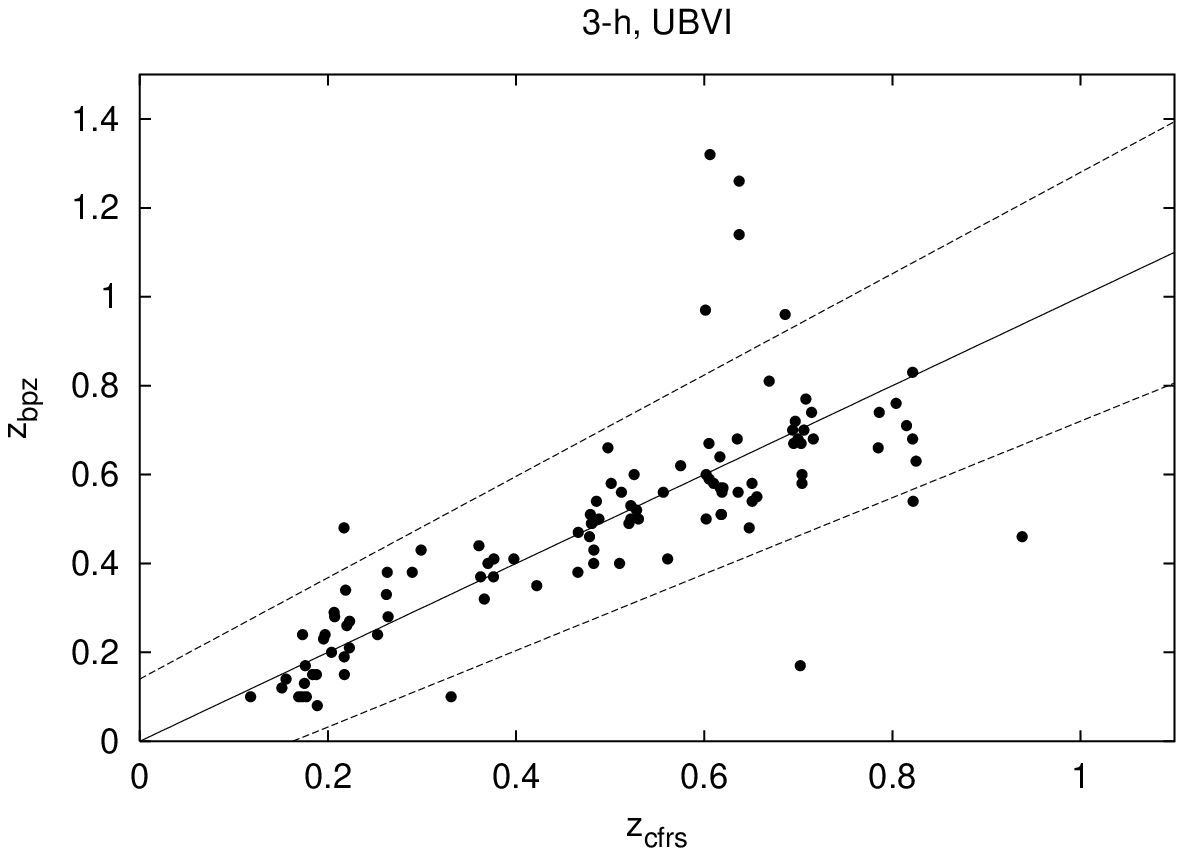,width=6.5cm,height=4.5cm}} 
 \subfigure[\label{photspecb}]{\psfig{file=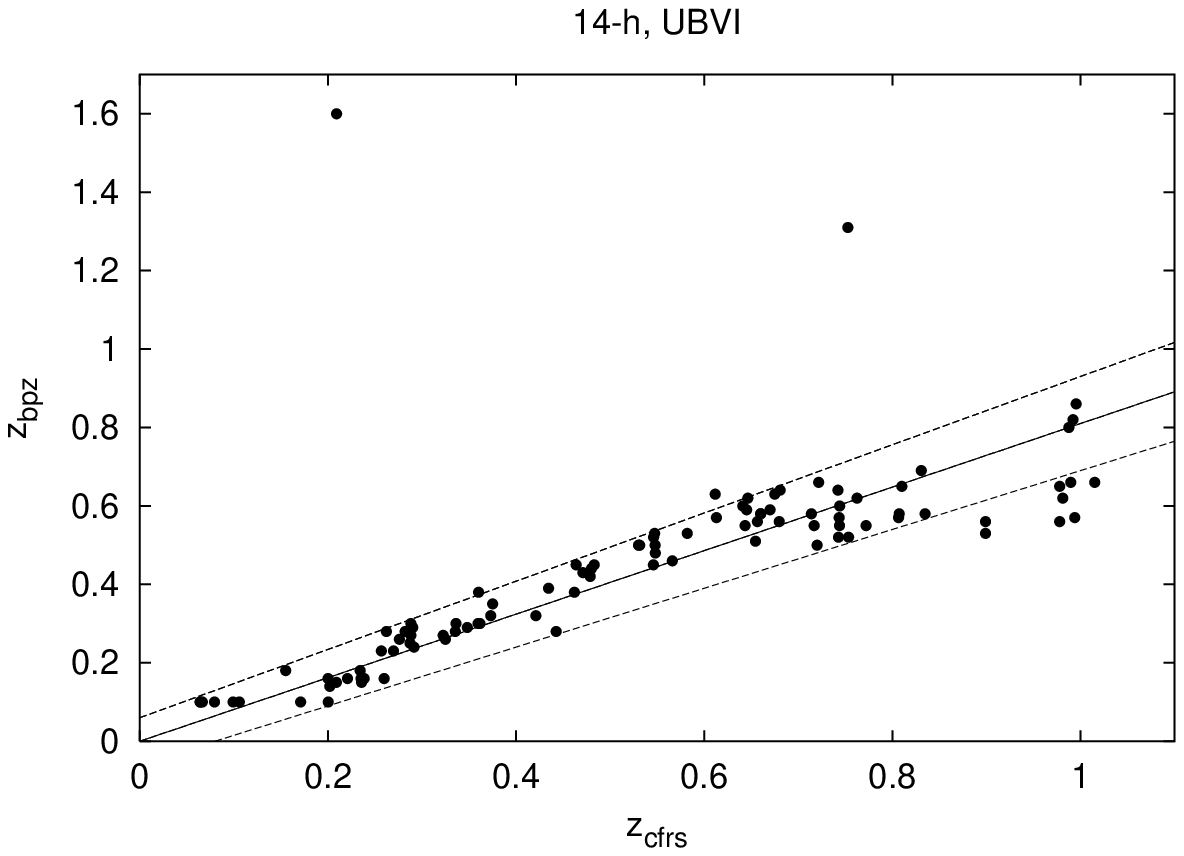,width=6.5cm,height=4.5cm}}
 \subfigure[\label{photspecc}]{\psfig{file=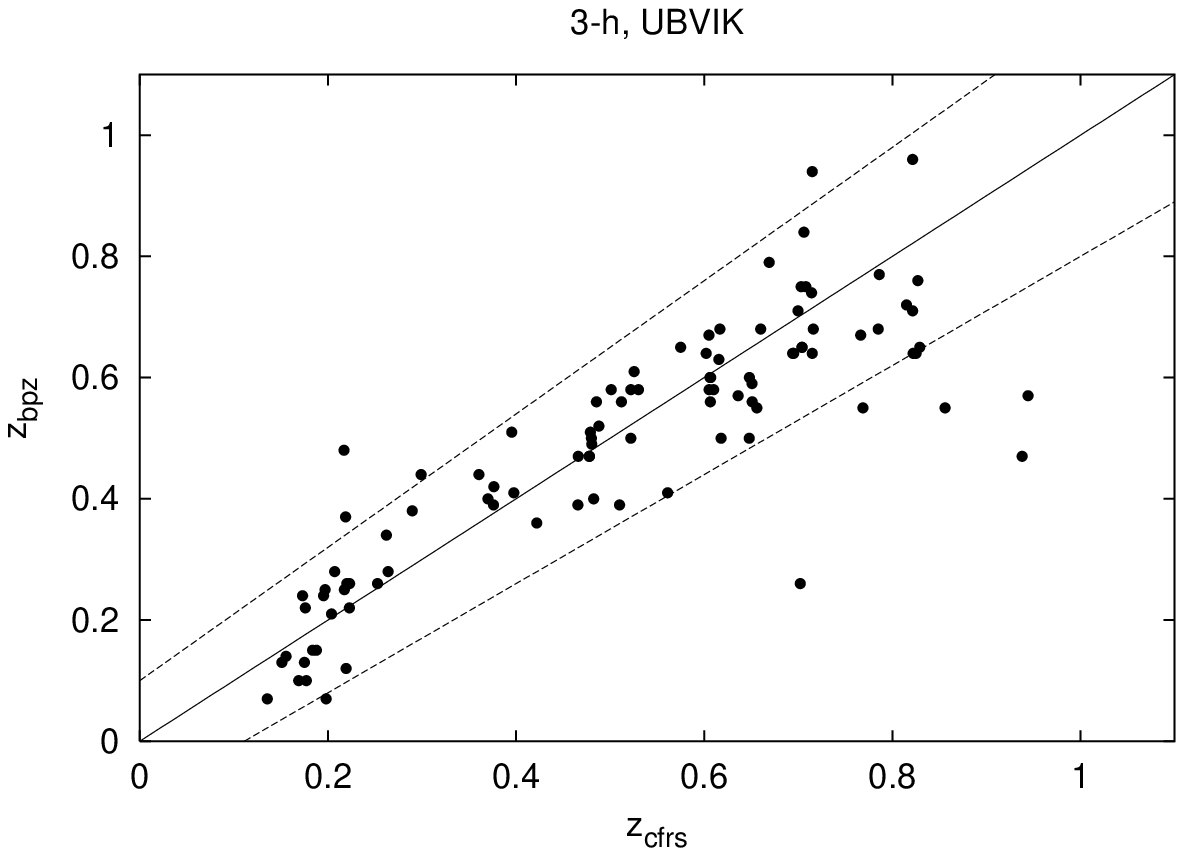,width=6.5cm,height=4.5cm}}
 \subfigure[\label{photspecd}]{\psfig{file=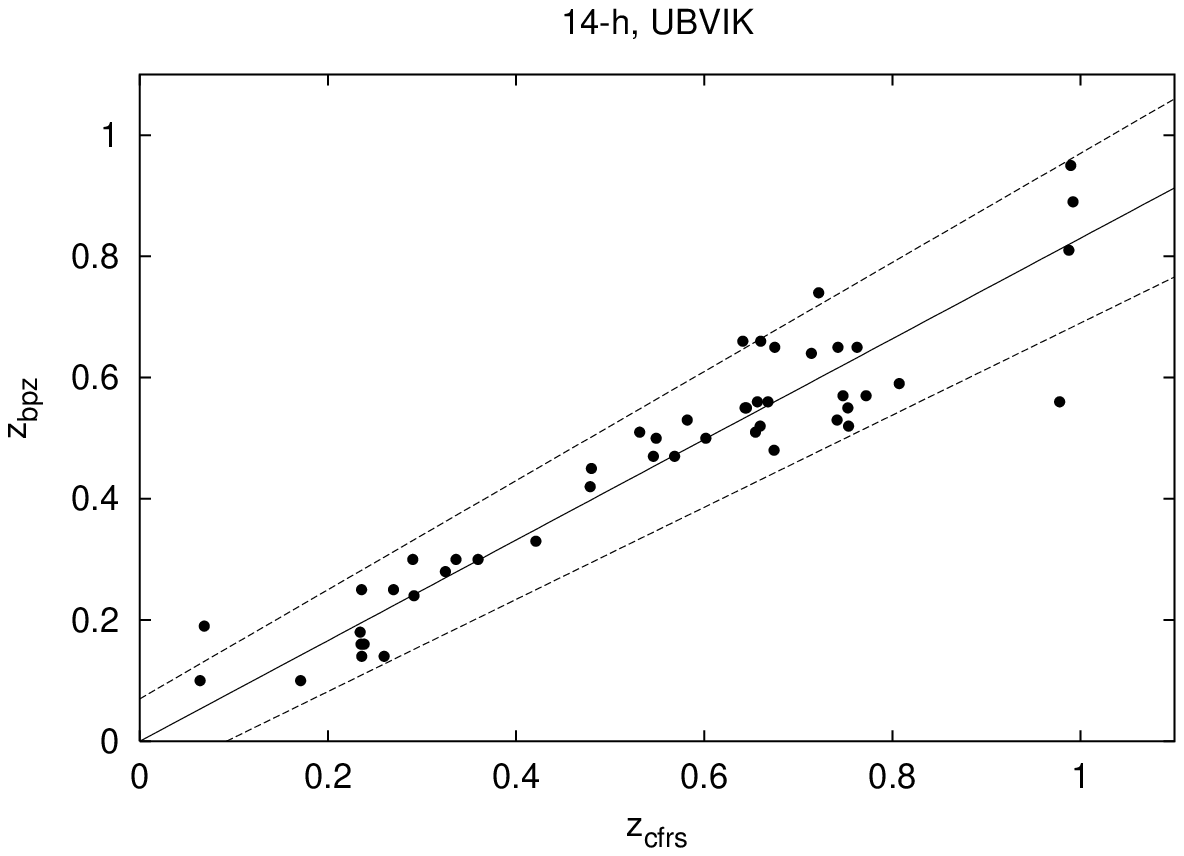,width=6.5cm,height=4.5cm}}
 \caption{\label{photspec}Photometric vs. spectroscopic redshift for
 the 3-h (fig~\ref{photspeca}, \ref{photspecc}) and 14-h
 (fig~\ref{photspecb}, \ref{photspecd}) fields.  The best fit
 gradients for 14-h field are 0.81 for $UBVI$ and 0.83 for $UBVIK$.
 The error lines shown are of the form $\sigma(1+z)$ where
 $\sigma=0.14,0.06,0.1,0.07$ for the sequence of plots (for the 14-h
 $UBVI$ plot this error ignores the two outliers, $\sigma=0.19$ if
 they are included).}
\end{figure*} 

This code has been tested by Gonzalez \& Maccarone (2002) and
Ben\'{i}tez (2000) and has proven to be highly successful, but we use
a different filter set and photometry from these studies and so it was
prudent to re-test the code for our specific needs.

Figure~\ref{photspec} shows comparison of the BPZ photometric redshift
estimates with spectroscopic measurements from the CFRS catalogue, for
those sources covered by both surveys.  Photometric estimates with low
reliability (quantified by an in-code statistic, $P_{\Delta z}<0.95$)
are removed from these plots leaving the most reliable estimates.  We
achieve reasonably good results both with and without the inclusion of
$K$ band photometry from the CFRS (only about half the objects here
have $K$ data), although the 14-h field suffers from a slight
systematic underestimation, which can be seen in the figures.  All BPZ
redshifts have been corrected for this effect in the main body of this
paper.  In general the scatter of the photometric redshifts is of the
order of $\sigma\sim0.1$ (see figure~\ref{photspec} for details).

Equivalent plots for the CFDF code can be found in Brodwin et
al. (2003) (their figure 2).  In comparison to BPZ the CFDF code
redshifts fair rather better when compared to the CFRS spectroscopic
sample, with fewer outliers and a smaller scatter ($\sigma\sim0.04$ to
$I_{AB}=22.5$, $\sigma\sim0.06$ to $I_{AB}=24$).  There are also no
systematic effects, as seen in the BPZ 14-h sample.

\subsection*{CFDF Photometric Redshift Estimation Code}

\begin{figure}
 \psfig{file=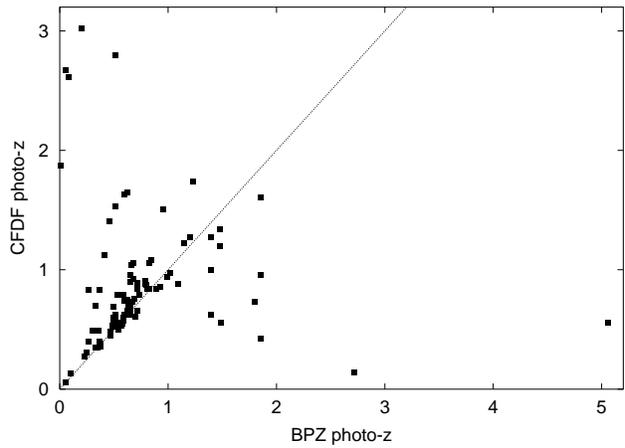,width=8.7cm,height=6.1cm}
 \caption{\label{zcomp}CFDF vs. BPZ photometric redshift estimates for
objects that have reliable estimates from both codes (see text for
details).  All possible stars and known QSOs have been excluded, as
have saturated objects with $I_{AB}<18.5$.}
\end{figure}

\begin{figure*}
 \subfigure[\label{Gubvi}]{\psfig{file=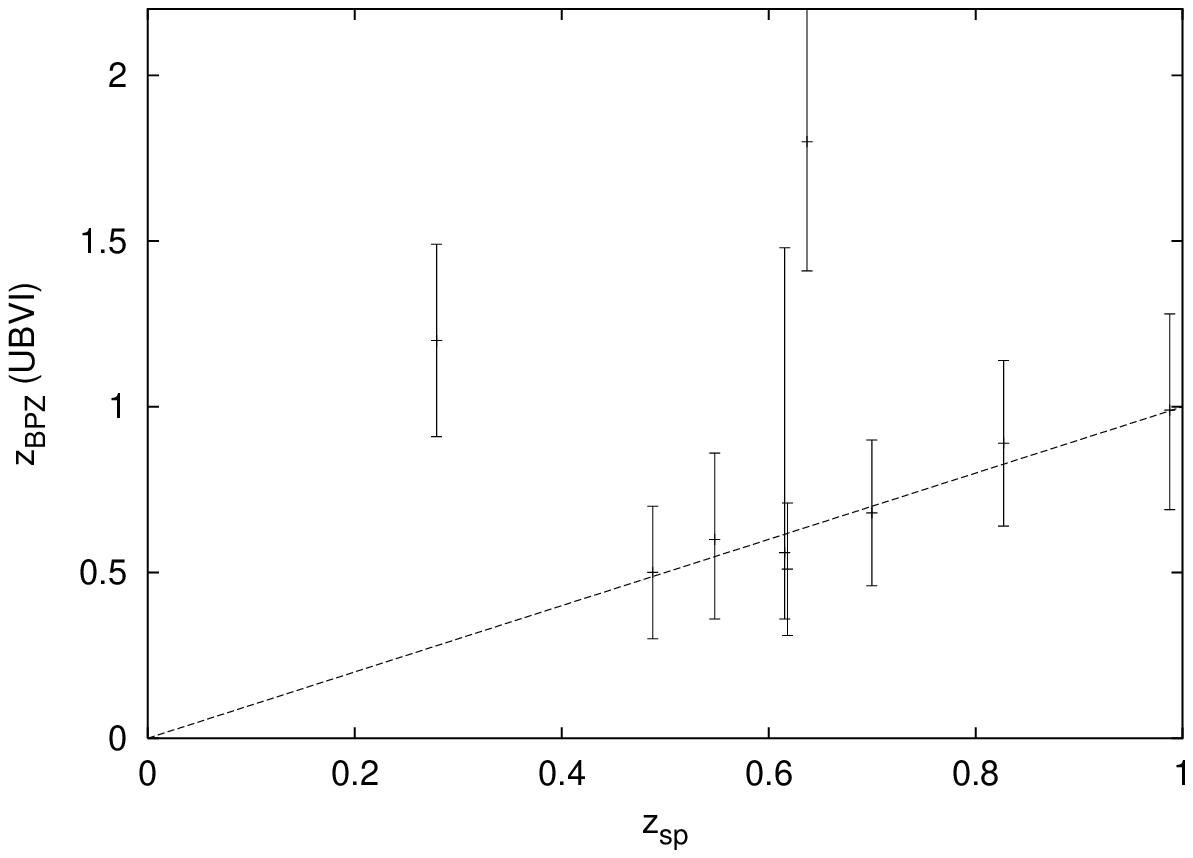,width=5.5cm,height=4cm}}
 \subfigure[\label{Gubvik}]{\psfig{file=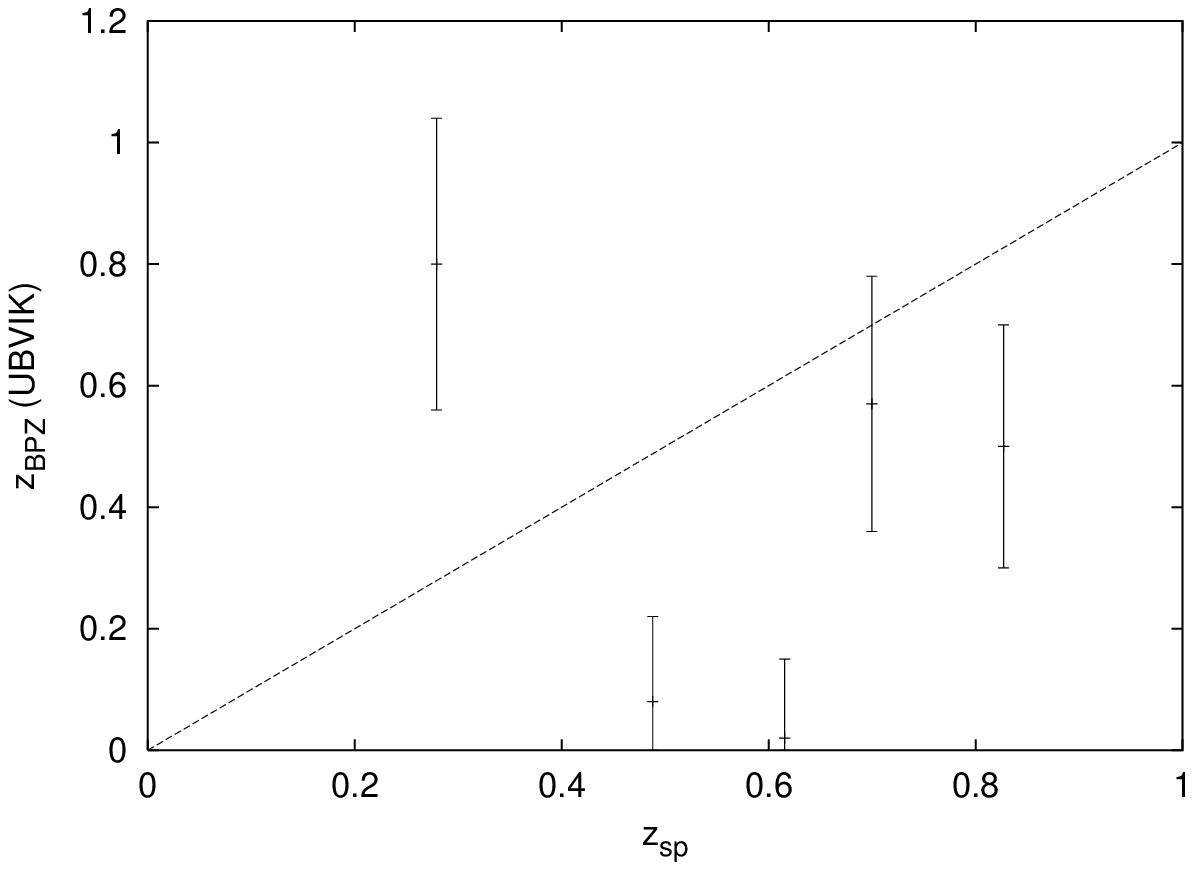,width=5.5cm,height=4cm}}
 \subfigure[\label{Gubvriz}]{\psfig{file=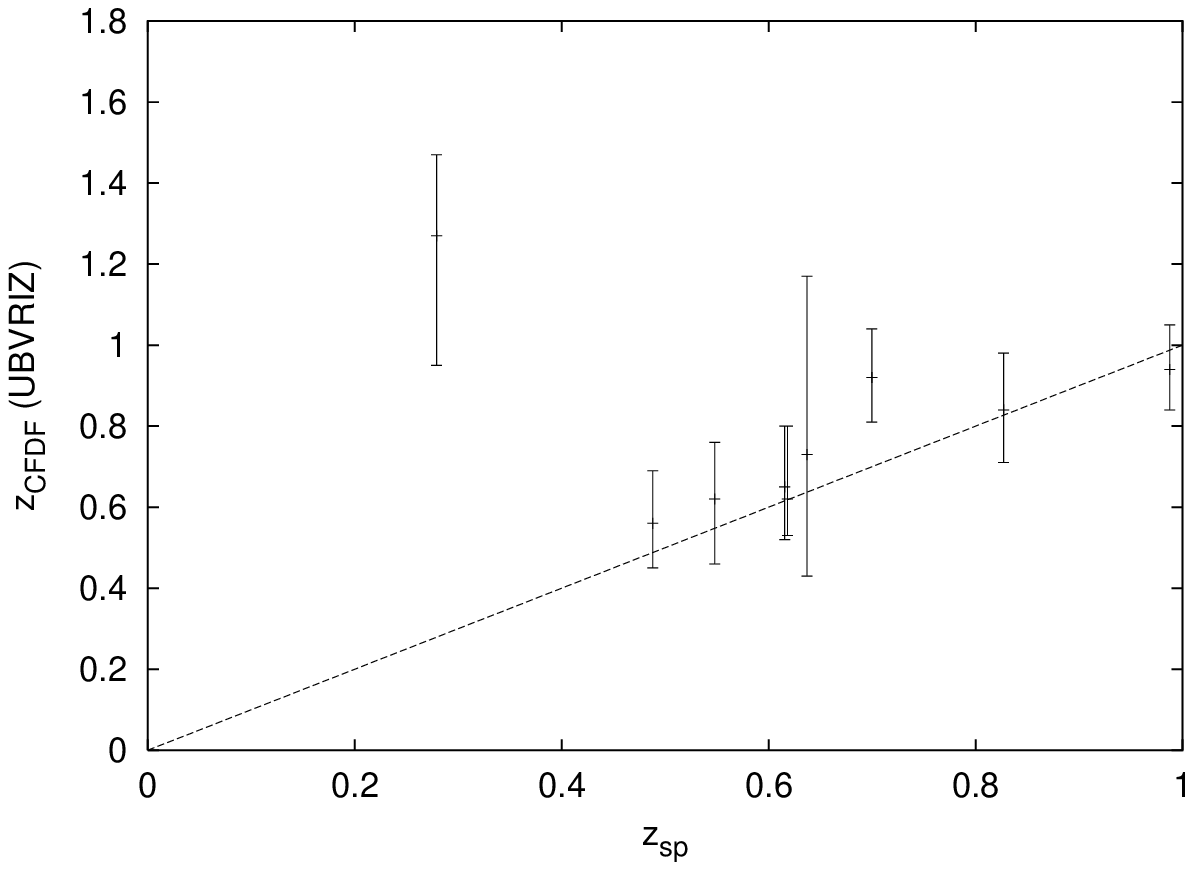,width=5.5cm,height=4cm}}
 \subfigure[\label{Qubvi}]{\psfig{file=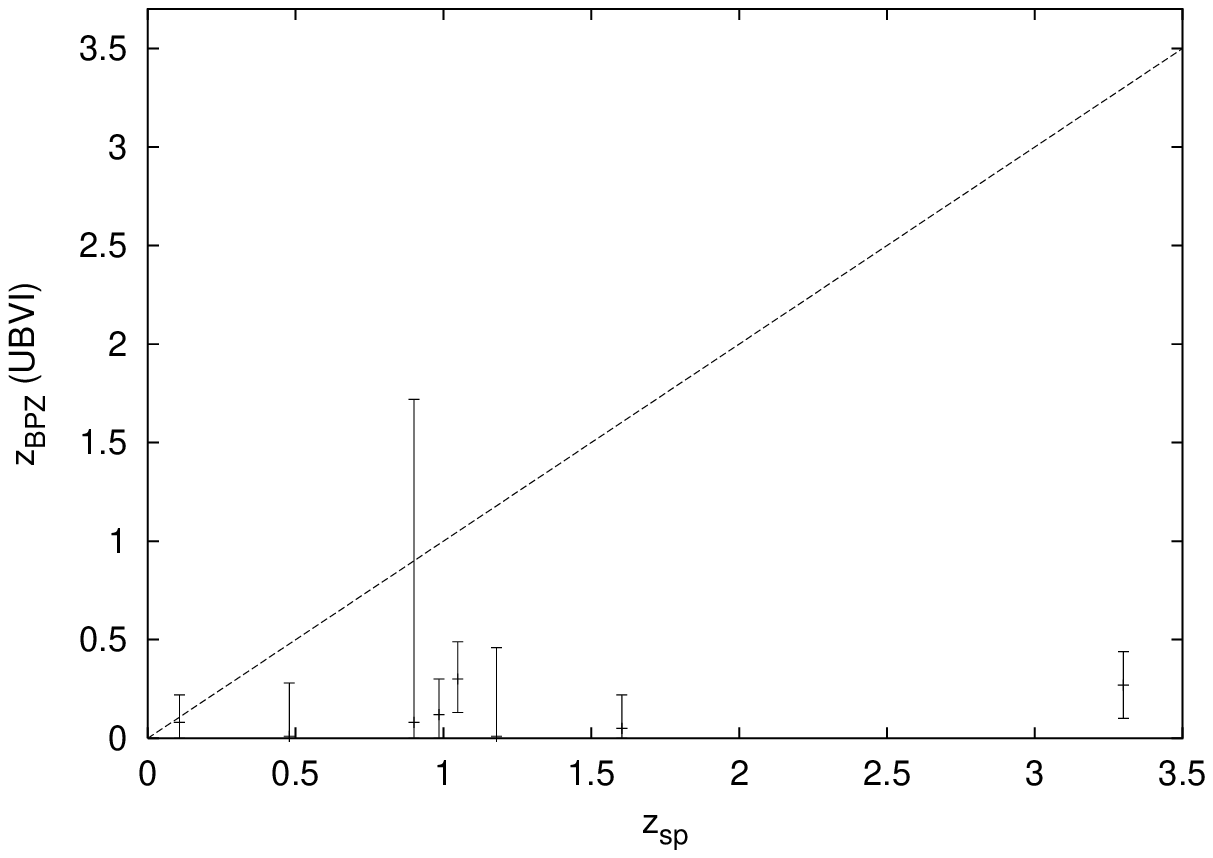,width=5.5cm,height=4cm}}
 \subfigure[\label{Qubvik}]{\psfig{file=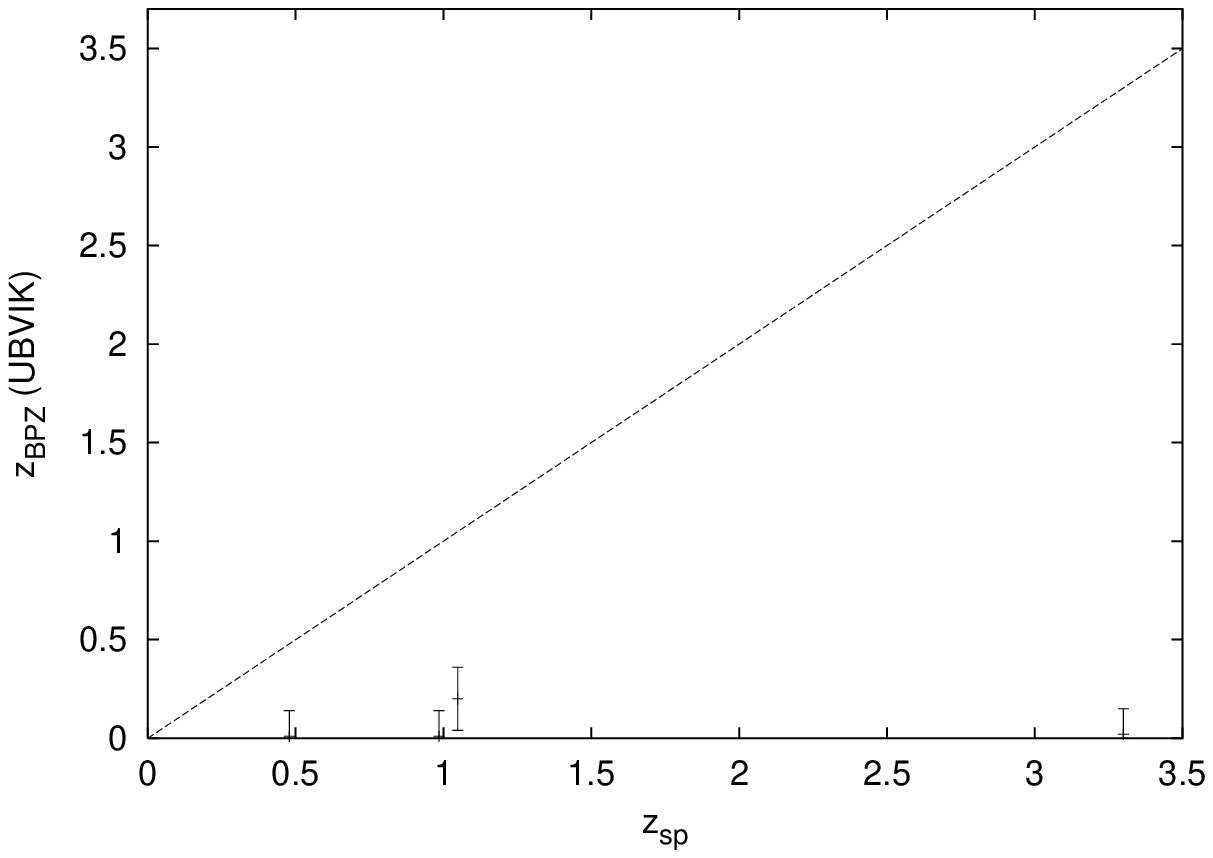,width=5.5cm,height=4cm}}
 \subfigure[\label{Qubvriz}]{\psfig{file=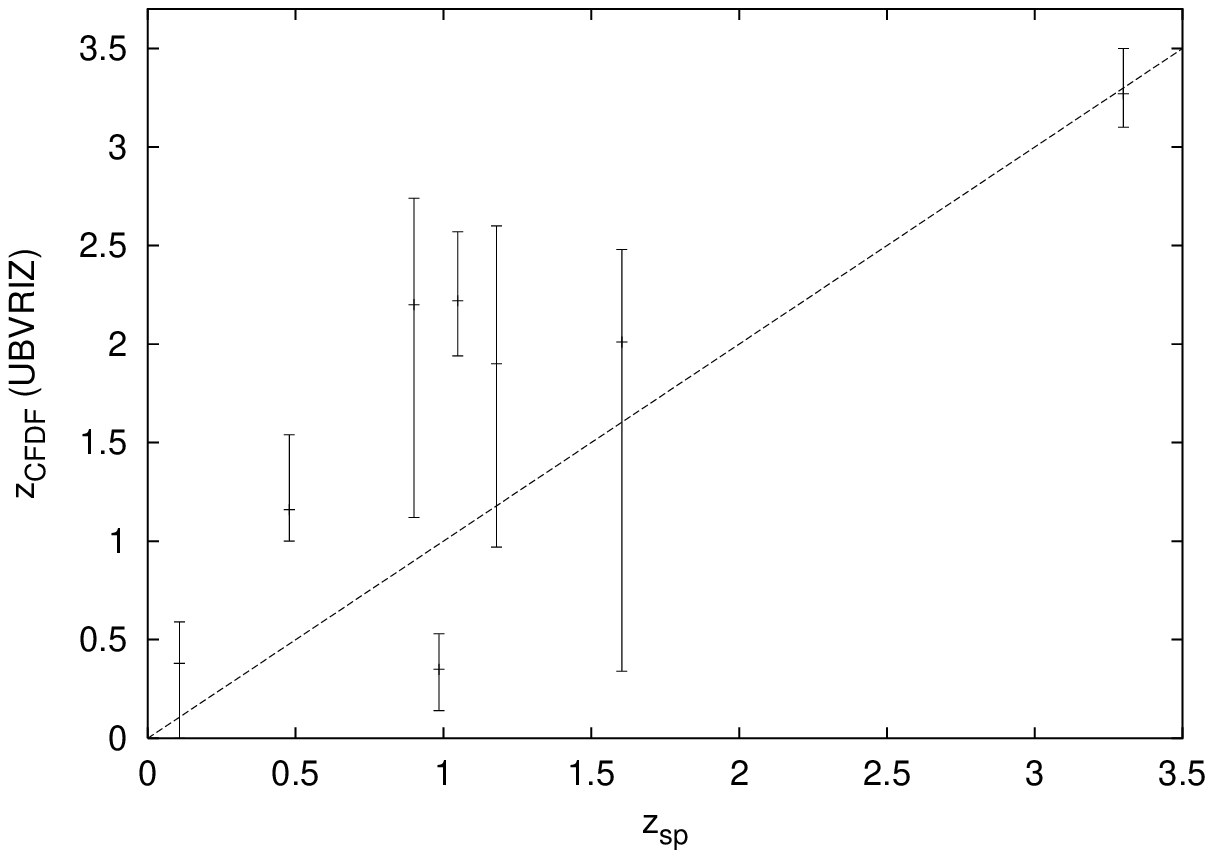,width=5.5cm,height=4cm}}
 \caption{\label{photoz}A series of plots to demonstrate the
 effectiveness of the two photo-z codes when compared to the handful
 of X-ray sources that have spectroscopic redshifts in our surveys.
 The top row (\ref{Gubvi}, \ref{Gubvik} \& \ref{Gubvriz}) shows the
 results for the non-QSO X-ray sources.  The bottom row (\ref{Qubvi},
 \ref{Qubvik} \& \ref{Qubvriz}) shows the results for the known QSOs.
 The left plots (\ref{Gubvi} \& \ref{Qubvi}) are BPZ results with
 $UBVI$ photometry, the middle (\ref{Gubvik} \& \ref{Qubvik}) are BPZ
 results for the few $UBVIK$ objects and the right plots
 (\ref{Gubvriz} \& \ref{Qubvriz}) show the results for the CFDF code
 with $UBVRIZ$ photometry.  The error bars in all cases are the
 $95~per~cent$ confidence limits around the best fit redshift.  The
 CFDF code is marginally more effective in our tests, and seems to
 handle QSOs more effectively, although still with larger error-bars
 than non-QSOs.  In general, for the CFRS (see main text and
 figure~\ref{photspec}), the addition of $K$ data appears to improve
 the BPZ results, however this effect is questionable for the 5 X-ray
 sources shown in figure~\ref{Gubvik}.}
\end{figure*}

The CFDF is currently extending beyond the original $UBVI$ survey to
include additional $R$ and $Z$ photometry.  These extra filters remove
some potential redshift degeneracies in certain galaxy templates and
so should provide more reliable photometric redshift estimates.  The
CFDF photometric redshift program is now underway with the full
$UBVRIZ$ photometry using a code developed by Mark Brodwin (Brodwin et
al. 2003, which includes a thorough analysis of its reliability).  As
an additional check on the original estimates we made with BPZ we
obtained photometric redshifts for our identified X-ray sources from
the CFDF photometric redshift program utilising these new catalogues.

Figure~\ref{zcomp} shows the comparison between the redshift estimates
made by the two codes for only those objects which had reliable
estimates as judged by both of the in-code measures.  Star like objects,
known QSOs and saturated objects ($I_{AB}<18.5$) (see catalogue
tables) are excluded because photometric redshifts are unreliable for
these objects.  In general the agreement is good, with $79~per~cent$
of objects agreeing to within a factor of 1.7.  The agreement is also
better for $z<1$, where the majority of objects lie (72/94 CFDF; 77/94
BPZ) and where the peak in the number density of intermediate
luminosity AGN is (Cowie et al. 2003).  Assuming the CFDF redshifts
are correct 7/72 $z<1$ objects are given poor redshifts by BPZ;
whereas assuming the BPZ redshifts are correct 13/77 $z<1$ objects are
given poor redshifts by the CFDF code.  

For the actual X-ray sources considered in this work only a handful
have spectroscopically measured redshifts, with half of these being
previously known QSOs.  Figure~\ref{photoz} shows the results of the
photometric redshift codes for all these objects.  There is a clear
problem in obtaining photometric redshifts for QSOs, both codes
struggling to pin them down with any accuracy.  However, for more
optically normal AGN both BPZ and the CFDF code cope quite well for
the most part.  The $95~per~cent$ confidence limits are slightly
better for the CFDF code however, and it also wins out over BPZ with
fewer unreliable redshifts in our X-ray sample.  Interestingly enough
the inclusion of $K$ band data to the BPZ code does not improve the
estimates, as we see in our tests of BPZ on the CFRS sources (above).
Although the sample here is small it actually appears to have an
adverse effect on the redshift estimations (figure~\ref{Gubvik})
rather than improving them as expected.

The CFDF code, being specifically designed for the objects used as IDs
in this survey, seems the logical choice for obtaining redshifts for
the X-ray sources.  This is especially true given that it also takes
full advantage of the more recent extension in the number of filters
for the CFDF.  This code does allow for the use of a Bayesian prior,
like BPZ, although none was used in obtaining these particular results
(instead the data itself is used to derive a prior for statistical
analysis of the full sample).  This may be seen as a slight
disadvantage, since priors have been shown to be effective in reducing
the number of catastrophic errors for individual galaxies (Ben\'{i}tez
2000).  However, the extra photometry used in the CFDF code should
compensate for this to some degree, and our tests and those in Brodwin
et al. (2003) show the CFDF code to be superior to BPZ in this
situation.


\begin{thebibliography}{}

\bibitem {} Alexander D. M., Brandt W. N., Hornschemeier A. E.,
Garmire G. P., Schneider D. P., Bauer F. E., Griffiths R. E., 2001,
AJ, 124, 1839
\bibitem {} Alexander D. M. et al., 2003, AJ, 125, 383
\bibitem {} Almaini O., Lawrence A., Boyle B. J., 1999, MNRAS, 305,
L59
\bibitem {} Baldi A., Molendi S., Comastri A., Fiore F., Matt G.,
Vignali C., 2002, ApJ, 564, 190
\bibitem {} Barcons X. et al., 2002, A\&A, 382, 522
\bibitem {} Barger A. J., Cowie L. L., Brandt W. N., Capak P., Garmire 
G. P., Hornschemeier A. E., Steffen A. T., Wehner E. H., 2002, AJ,
124, 1839
\bibitem {} Barger et al., 2003, AJ, 126, 632
\bibitem {} Ben\'{i}tez N., 2000, ApJ, 536, 571
\bibitem {} Brodwin et al. 2003, ApJ, submitted (astro-ph/0310038)
\bibitem {} Chapman S. C. et al., 2003, ApJ, 585, 57
\bibitem {} Coleman G. D., Wu C-C., Weedman D. W., 1980, ApJS, 43, 393
\bibitem {} Cowie L. L., Barger A. J., Bautz M. W., Brandt W. N.,
Garmire G. P., 2003, ApJ, 584, L57
\bibitem {} Csabai I., et al., 2003, AJ, 125, 580
\bibitem {} Downes A. J. B., Peacock J. A., Savage A., Carrie D. R.,
1986, MNRAS, 218, 31
\bibitem {} Fabian A. C., Barcons X., 1992, ARA\&A, 30, 429
\bibitem {} Fontana A., D'Odorico S., Poli F., Giallongo E., Arnouts
S., Cristiani S., Moorwood A., Saracco P., 2000, AJ, 120, 2206
\bibitem {} Freeman P. E., Kashyap V., Rosner R., Lamb D. Q., 2002,
ApJS, 138, 185
\bibitem {} Gandhi P., Crawford C. S., Fabian A. C., Johnstone R. M.,
2003, MNRAS in press (astro-ph/0310772)
\bibitem {} Giacconi R. et al., 2002, ApJS, 139, 369
\bibitem {} Gilli R., Salvati M., Hasinger G., 2001, A\&A, 366, 407
\bibitem {} Gonzalez A. H., Maccarone T. J., 2002, ApJ, 581, 155
\bibitem {} Hammer F., Crampton D., Le F\`{e}vre O., Lilly S. J.,
1995, ApJ, 455, 88
\bibitem {} Hasinger G., Burg R., Giacconi R., Schmidt M., Tr\"{u}mper 
J., Zamorani G., 1998, A\&A, 329, 482
\bibitem {} Hasinger G. et al., 2001, A\&A, 365, L45
\bibitem {} Hornschemeier A. E., Bauer F. E., Alexander D. M., Brandt
W. N., Sargent W. L., Vignali C., Garmire G. P., Schneider D. P.,
2003, AN, 324, 12
\bibitem {} Kashikawa N. et al., 2003, AJ, 125, 53
\bibitem {} Kinney A. L., Calzetti D., Bohlin R. C., McQuade K.,
Storchi-Bergmann T., Schmitt H. R., 1996, ApJ, 467, 38
\bibitem {} Lilly S. J., Le F\`{e}vre O., Crampton D., Hammer F., 
Tresse L., 1995a, ApJ, 455, 50
\bibitem {} Lilly S. J., Hammer F., Le F\`{e}vre O., Crampton D.,
1995b, ApJ, 455, 75
\bibitem {} Magorrian et al., 1998, AJ, 155, 2285
\bibitem {} Mainieri V., Bergeron J., Hasinger G., Lehmann I., Rosati
P., Schmidt M., Szokoly G., Della Ceca R., 2002, A\&A, 393, 425
\bibitem {} McCammon D., Sanders W. T., 1990, ARA\&A, 28, 657
\bibitem {} McCracken H. J., Le F\`{e}vre O., Brodwin M., Foucaud S.,
Lilly S. J., Crampton D., Mellier Y., 2001, A\&A, 376, 756
\bibitem {} M$^{c}$Hardy I. M. et al., 1998, MNRAS, 295, 641
\bibitem {} M$^{c}$Hardy I. M. et al., 2003, MNRAS, 342, 802
\bibitem {} Miyaji T., Griffiths R. E., 2001, Recent results of
XMM-Newton and Chandra, XXXVIth Rencontres de Moriond, XXIst Moriond
Astrophysics Meeting, eds D.M. Neumann \& J.T.T. Van, 65M
\bibitem {} Moretti A., Campana S., Lazzati D., Tagliaferri G., 2003,
ApJ, 588, 696
\bibitem {} Nandra et al., 2004, in preparation
\bibitem {} Page M. J. et al., 2003, AN, 324, 101
\bibitem {} Severgnini P. et al., 2000, A\&A, 360, 457
\bibitem {} Steffen A. T., Barger A. J., Cowie L. L., Mushotzky R. F., 
Yang Y., 2003, ApJ, 596, L23
\bibitem {} Ueda Y., Akiyama M., Ohta K., Miyaji T., 2003, ApJ, 598, 886
\bibitem {} Waskett et al., 2003, MNRAS, 341, 1217 

\end{thebibliography}
\end{document}